\documentclass[aps,physrev,reprint,superscriptaddress,longbibliography]{revtex4-2}
\usepackage{amsmath,amssymb,bm}
\usepackage{graphicx}
\usepackage{CJKutf8}

\begin{document}
	    \begin{CJK*}{UTF8}{gbsn}
		
		\title{Network topology imposes spectral constraints on collective nonequilibrium responses}
		
		\author{Dingjie Wu (吴丁杰)}
		\affiliation{
			State Key Laboratory for Strength and Vibration of Mechanical Structures,
			Xi'an Jiaotong University, Xi'an 710049, China
		}
		\affiliation{
			School of Aerospace Engineering, Xi'an Jiaotong University, Xi'an 710049, China
		}
		
		\author{Yaozu Wang}
		\affiliation{
			College of Intelligence Science and Technology, National University of Defense Technology, Changsha 410073, China
		}
		
		\author{Rong Wang}
		\email[Contact author: ]{tobous@mail.xjtu.edu.cn}
		\affiliation{
			State Key Laboratory for Strength and Vibration of Mechanical Structures,
			Xi'an Jiaotong University, Xi'an 710049, China
		}
		\affiliation{
			School of Aerospace Engineering, Xi'an Jiaotong University, Xi'an 710049, China
		}
		
		\author{Ying Wu}
		\email[Contact author: ]{wying36@xjtu.edu.cn}
		\affiliation{
			State Key Laboratory for Strength and Vibration of Mechanical Structures,
			Xi'an Jiaotong University, Xi'an 710049, China
		}
		\affiliation{
			School of Aerospace Engineering, Xi'an Jiaotong University, Xi'an 710049, China
		}
		
		\date{\today}
		
		\begin{abstract}
			Collective nonequilibrium dynamics in Langevin systems depend
			on both network topology and stochastic driving.
			Understanding the role of topology in responses to stochastic
			driving and in the emergence of nonequilibrium is essential.
			Starting from the principle of microscopic reversibility and linear
			Langevin dynamics on networks, we derive a mode decomposition for
			both the asymmetry of short-time cross-correlations and heat currents
			at steady state.
			This decomposition reveals two sources of nonequilibrium---fluctuations
			of individual modes and correlated fluctuations between distinct modes.
			For reciprocal networks with independent local driving, we derive
			two nonequilibrium response operators that yield explicit expressions
			for the global asymmetry intensity and the entropy production rate. These operators are weighted Laplacians independent of stochastic driving. Their spectra determine attainable upper and lower bounds on nonequilibrium at fixed driving heterogeneity.
			These bounds are exact for asymmetry intensity at arbitrary
			deviations from equilibrium and hold to leading order near
			equilibrium for entropy production rate. Finally, we find that the hierarchical modular organization of human structural connectomes broadens the brain’s nonequilibrium operating range, with resting-state brain activity supporting this topological advantage
			for sustaining a lower nonequilibrium level. Our study provides a basis for understanding biological network organization and designing networks with targeted nonequilibrium operating points.
		\end{abstract}
		\maketitle
	    \end{CJK*}
	
		\section{Introduction}
        Many complex systems operate out of equilibrium and can be
        modeled using Langevin dynamics~\cite{Seifert2012,OByrne2022,Gnesotto2018,Battle2016}.
        In such systems, deterministic interactions among constituent
        units and stochastic driving jointly determine the collective
        dynamics, e.g., in gene regulatory networks with intrinsic biochemical
        noise~\cite{Paulsson2005,Aslyamov2026}, active matter driven by stochastic active
        forces~\cite{Fodor2016,Flenner2016} and neural systems receiving noisy synaptic inputs~\cite{Destexhe2001,Chance2002}. Stochastic driving excites local fluctuations that propagate
        across the system through interactions among units and can generate
        collective nonequilibrium dynamics~\cite{Gladrow2017,Mura2018,Gradziuk2019}. At steady state, these dynamics are characterized by the entropy
        production rate~\cite{Godreche2019}, persistent heat
        currents~\cite{Ciliberto2013} and time-reversal
        asymmetry~\cite{Ohga2023}.
		
		Structurally, different network topologies constrain how local
		fluctuations coalesce into collective
		fluctuations~\cite{Ren2010,Lizier2023,NartalloKaluarachchi2024}.
		In recent decades, tuning network topology in physical and biological systems has provided an effective control knob for modulating collective nonequilibrium behavior by harnessing stochastic driving to amplify or suppress fluctuations~\cite{Korniss2007,Yanagita2012,Hennequin2012,Ronellenfitsch2018},
		regulate dissipation~\cite{Corson2010,Herpich2018,Zhang2020,Freitas2020,Meibohm2024} and enhance thermodynamic efficiency~\cite{Vroylandt2017,Mamede2023,Filho2023}. These studies show that network topology can be tuned to achieve specific responses to stochastic driving and thereby constrain the system's nonequilibrium operating range. However, the mechanisms by which network topology modulates responses to stochastic driving and thereby constrains collective nonequilibrium dynamics remain unclear. Understanding these mechanisms would guide the optimal design of artificial networks and reveal structural advantages underlying nonequilibrium operation in biological networks~\cite{Lan2012,Zheng2024,NartalloKaluarachchi2026}.
		
	    Dynamically, intricate network topologies govern deterministic network dynamics that support a set of distinct intrinsic modes~\cite{Newman2003,Boccaletti2006,Pecora1998,Nakao2010}, while collective fluctuations can stem from the joint excitation of multiple modes by driving~\cite{Hunt2010,Tyloo2018,Hindes2023}. Interestingly, statistically correlated fluctuations of distinct modes have been linked to diverse nonequilibrium
	    phenomena~\cite{Fyodorov2025,Asllani2018,Gladrow2017,Argun2017,Falasco2015}. Such correlations can arise from the nonorthogonality of modes in nonnormal network dynamics~\cite{Fyodorov2025,Asllani2018}. Even when the modes are orthogonal, spatially correlated or heterogeneous stochastic driving can correlate the excitation of distinct modes, as demonstrated by stochastic active forces in cytoskeletal networks~\cite{Gladrow2017} and anisotropic noise in colloidal Brownian gyrators~\cite{Argun2017}. When two correlated modes relax at different rates, their lagged cross-correlations can decay differently depending on which mode
	    is observed later, revealing time-reversal asymmetry in the collective dynamics~\cite{Crisanti2012}. These observations suggest that network topology may constrain collective nonequilibrium dynamics through the correlated responses of distinct modes to stochastic driving.
	    
		In this work, based on the principle of microscopic reversibility and linear Langevin dynamics on a network, we first identify two sources of collective nonequilibrium dynamics: time-reversal asymmetry in (i) \emph{fluctuations of individual modes} and (ii) \emph{correlated fluctuations between two modes}. This mode representation provides an additive decomposition of the asymmetry of short-time cross-correlations and heat currents at steady state. For reciprocal networks with independent local driving, we derive two \emph{nonequilibrium response operators} for the global
		asymmetry intensity \(\omega\) and the entropy production rate \(\sigma\). Both operators are weighted Laplacians determined by the network dynamics and independent of the imposed driving. At fixed heterogeneous driving strength, their spectra give exact
		upper and lower bounds on \(\omega\) at arbitrary deviations from
		equilibrium and leading-order bounds on \(\sigma\) near equilibrium.
		The generalized spectrum further determines upper and lower bounds
		on \(\omega/\sigma\) near equilibrium.
		The corresponding eigenvectors identify stochastic driving
		patterns that attain these bounds.
		
		We apply this theoretical framework to synthetic hierarchical
		modular networks and human structural connectomes reconstructed
		from diffusion-weighted imaging (DWI) data from \(993\) participants.
		Hierarchical organization and multiscale connection weights
		broaden the nonequilibrium operating range, revealing a
		nonequilibrium advantage of hierarchical modular networks
		that are widespread in nature.
		Furthermore, using resting-state functional magnetic resonance
		imaging (fMRI) data, we find that the specific stochastic driving
		patterns at rest help maintain brain activity at a low
		nonequilibrium level, providing a possible dynamical explanation for the low
		nonequilibrium level reported in resting-state brain
		activity~\cite{Kringelbach2023,Deco2023}.
		Our study quantifies how network topology constrains collective
		nonequilibrium responses to stochastic driving, providing a basis
		for understanding biological network organization and designing
		networks with desired operating points and ranges.
		
		\section{Topological constraints on collective nonequilibrium responses}
		
		\subsection{Linear Langevin dynamics on networks}
		We consider a network of \(N\) nodes whose dynamics obey the overdamped linear Langevin equation
		\begin{equation}
			\dot{\mathbf{x}}(t)=-\mathbf{A}\mathbf{x}(t)+\boldsymbol{\eta}(t),
			\label{eq:langevin}
		\end{equation}
		where \(\mathbf x(t)\) collects the node variables. The matrix \(\mathbf A\) represents the interaction network, e.g., interparticle interactions in active matter~\cite{Mura2018} or synaptic coupling among neurons~\cite{Hennequin2014}.
		We assume that all eigenvalues of \(\mathbf A\) have positive real
		parts to ensure relaxation to a unique steady state.
		Such linear dynamics also provide a local approximation to nonlinear
		dynamics near a stable fixed point, with \(\mathbf x(t)\) representing deviations from that point and \(-\mathbf A\) denoting the Jacobian of the deterministic dynamics evaluated there~\cite{Kwon2005,McKane2005,Thomas2012}. The stochastic driving \(\boldsymbol{\eta}(t)\) is Gaussian and white,
		with zero mean and covariance
		\(\langle\boldsymbol{\eta}(t)\boldsymbol{\eta}^{\mathrm T}(t')\rangle
		=2\mathbf D\,\delta(t-t')\),
		where \(\mathbf D\) is a symmetric positive definite diffusion matrix.
		Its diagonal entries \(D_{ii}\) specify the local driving intensities,
		while its off-diagonal entries \(D_{ij}\) encode spatial correlations
		between the driving at different nodes.
		
		\subsection{Mode decomposition of collective nonequilibrium dynamics}
		The lagged covariance at steady state is defined as
		\(\mathbf C_x(\tau)=\langle\mathbf x(t+\tau)\mathbf x^{\mathrm T}(t)\rangle_0\),
		where \(\langle\cdot\rangle_0\) denotes an average at steady state.
		For the linear dynamics in Eq.~\eqref{eq:langevin},
		\(\mathbf C_x(\tau)=\mathbf G(\tau)\mathbf C_0\) for \(\tau\geq0\),
		where \(\mathbf G(\tau)=e^{-\mathbf A\tau}\) is the network propagator
		and \(\mathbf C_0=\mathbf C_x(0)\) is the zero-lag covariance.
		For variables even under time reversal, the principle of microscopic
		reversibility requires
		\(\mathbf C_x(\tau)=\mathbf C_x^{\mathrm T}(\tau)\)
		for all \(\tau>0\) at equilibrium~\cite{Onsager1931,Casimir1945}.
		At a nonequilibrium steady state, this symmetry is violated
		for at least one \(\tau>0\),
		\begin{equation}
			\mathbf C_x(\tau)\neq\mathbf C_x^{\mathrm T}(\tau).
			\label{eq:covariance-reversibility}
		\end{equation}
		The zero-lag covariance \(\mathbf C_0\) satisfies the Lyapunov
		equation~\cite{Wang1945,Lax1960,Ferreira2025},
		\begin{equation}
			\mathbf A\mathbf C_0+\mathbf C_0\mathbf A^{\mathrm T}
			=2\mathbf D.
			\label{eq:lyapunov}
		\end{equation}
		Substituting its integral solution into
		\(\mathbf C_x(\tau)=\mathbf G(\tau)\mathbf C_0\),
		Eq.~\eqref{eq:covariance-reversibility} becomes
		\begin{equation}
			\int_0^\infty
			e^{-\mathbf A(t+\tau)}
			\mathbf D
			e^{-\mathbf A^{\mathrm T}t}\,\mathrm dt
			\!\neq\!
			\int_0^\infty
			e^{-\mathbf A t}
			\mathbf D
			e^{-\mathbf A^{\mathrm T}(t+\tau)}\,\mathrm dt.
			\label{eq:integral-reversibility}
		\end{equation}
		For a diagonalizable
		\(\mathbf A=\mathbf U\boldsymbol{\Lambda}\mathbf U^{-1}\),
		\(\mathbf U=(\mathbf U_1,\ldots,\mathbf U_N)\) collects
		the right eigenvectors and
		\(\boldsymbol{\Lambda}=\operatorname{diag}(\lambda_1,\ldots,\lambda_N)\)
		contains the corresponding eigenvalues.
		The mode coordinates \(\mathbf y(t)=\mathbf U^{-1}\mathbf x(t)\)
		obey
		\begin{equation}
			\dot{\mathbf y}(t)
			=-\boldsymbol{\Lambda}\mathbf y(t)+\boldsymbol{\xi}(t),
			\label{eq:mode-dynamics}
		\end{equation}
		where \(\boldsymbol{\xi}(t)=\mathbf U^{-1}\boldsymbol{\eta}(t)\)
		satisfies
		\(\langle\boldsymbol{\xi}(t)\boldsymbol{\xi}^{\dagger}(t')\rangle
		=2\boldsymbol{\Gamma}\delta(t-t')\).
		Here, \(\boldsymbol{\Gamma}=\mathbf U^{-1}\mathbf D\mathbf U^{-\dagger}\)
		is the mode diffusion matrix and \(\dagger\) denotes conjugate transpose.
		Although the deterministic dynamics are decoupled in these coordinates,
		the transformed driving can remain correlated between modes
		through the off-diagonal entries of \(\boldsymbol{\Gamma}\).
		The lagged covariance of mode fluctuations is
		\(\mathbf C_y(\tau)=\langle\mathbf y(t+\tau)\mathbf y^{\dagger}(t)\rangle_0
		=\mathbf U^{-1}\mathbf C_x(\tau)\mathbf U^{-\dagger}\).
		Since \(\mathbf U\) is invertible,
		\(\mathbf C_x(\tau)\) is symmetric if and only if
		\(\mathbf C_y(\tau)\) is Hermitian, Eq.~\eqref{eq:integral-reversibility} is equivalent to
		\begin{equation}
			\int_0^\infty
			e^{-\boldsymbol{\Lambda}(t+\tau)}
			\boldsymbol{\Gamma}
			e^{-\boldsymbol{\Lambda}^{\dagger}t}\,\mathrm dt
			\neq
			\int_0^\infty
			e^{-\boldsymbol{\Lambda}t}
			\boldsymbol{\Gamma}
			e^{-\boldsymbol{\Lambda}^{\dagger}(t+\tau)}\,\mathrm dt.
			\label{eq:modal-integral-reversibility}
		\end{equation}
		The diagonal entries
		\(C_{y_i y_i}(\tau)
		=(\Gamma_{ii}/\operatorname{Re}\lambda_i)e^{-\lambda_i\tau}\)
		characterize \emph{fluctuations of individual modes}.
		The off-diagonal entries
		\(C_{y_i y_j}(\tau)
		=2\Gamma_{ij}e^{-\lambda_i\tau}/(\lambda_i+\lambda_j^*)\)
		characterize \emph{correlated fluctuations between distinct modes}.
		Comparing these covariances under the two temporal orderings gives
		\begin{equation}
			\frac{C_{y_i y_i}(\tau)}
			{C_{y_i y_i}^{*}(\tau)}
			=e^{-2\mathrm{i}\operatorname{Im}\lambda_i\,\tau},
			\qquad
			\frac{C_{y_i y_j}(\tau)}
			{C_{y_j y_i}^{*}(\tau)}
			=e^{(\lambda_j^*-\lambda_i)\tau},
			\label{eq:modal-correlation-ratio}
		\end{equation}
		where \(*\) denotes complex conjugation and the second ratio
		is defined for \(i\neq j\) with \(\Gamma_{ij}\neq0\).
		At equilibrium, both ratios must equal unity for all \(\tau>0\).
		Fluctuations of an individual mode contribute to time-reversal
		asymmetry when \(\operatorname{Im}\lambda_i\neq0\).
		In this case, the covariance acquires opposite phases under
		the two temporal orderings.
		This contribution requires complex eigenvalues and is therefore
		possible only in nonreciprocal networks with
		\(\mathbf A\neq\mathbf A^{\mathrm T}\)~\cite{NartalloKaluarachchi2024}.
		Correlated fluctuations between distinct modes contribute to
		time-reversal asymmetry when both \(\Gamma_{ij}\neq0\) and
		\(\lambda_i\neq\lambda_j^*\).
		The first condition ensures nonzero correlations between the modes,
		while the second implies that these correlations decay at different
		rates or acquire different phases under the two temporal orderings.
		A nonzero \(\Gamma_{ij}\) can arise from the nonorthogonality of modes
		(\(\mathbf U_k^\dagger\mathbf U_\ell\neq0\) for some \(k\neq\ell\))
		in nonnormal networks~\cite{Fyodorov2025,Asllani2018}
		or from spatially structured stochastic driving
		(\(\mathbf D\not\propto\mathbf I\))~\cite{Gladrow2017,Argun2017},
		both encoded in
		\(\boldsymbol{\Gamma}=\mathbf U^{-1}\mathbf D\mathbf U^{-\dagger}\).
		Further discussion and examples are provided in the
		Supplemental Material (SM)~\cite{SupplementalMaterial}.
	
		This mode representation decomposes nonequilibrium observables into
		additive contributions from fluctuations of individual modes and
		correlated fluctuations between modes.
		We first consider the asymmetry of short-time
		cross-correlations~\cite{Ohga2023,Gu2024},
		\begin{equation}
			\boldsymbol{\Omega}
			=
			\lim_{\tau\to0^+}
			\frac{\mathbf C_x(\tau)-\mathbf C_x^{\mathrm T}(\tau)}{2\tau},
			\label{eq:short-time-asymmetry}
		\end{equation}
		where \(\Omega_{mn}
		=\langle x_m(\mathbf A\mathbf x)_n
		-x_n(\mathbf A\mathbf x)_m\rangle_0/2\)
		is the mean signed area swept out per unit time by the joint
		trajectory in the \((x_n,x_m)\) plane.
		It can be estimated directly from short-lag covariances
		in experimental multivariate time series.
		Combining the short-time limit of the covariance
		above with the Lyapunov equation yields the spectral form
		\begin{equation}
			\boldsymbol{\Omega}
			=
			\sum_{i,j=1}^{N}
			\frac{\lambda_j^*-\lambda_i}
			{\lambda_i+\lambda_j^*}
			\Gamma_{ij}
			\mathbf U_i\mathbf U_j^\dagger.
			\label{eq:modal-time-asymmetry}
		\end{equation}
		The terms with \(i=j\) describe contributions from fluctuations
		of individual modes, whereas those with \(i\neq j\) describe
		contributions from correlated fluctuations between distinct modes.
		The factors \(\Gamma_{ij}\) specify the covariance of the driving
		in mode coordinates, while the eigenvalues determine how the
		resulting fluctuations contribute to time-reversal asymmetry.
		An analogous decomposition holds for the asymmetry of lagged
		cross-correlations \(\mathbf C_x(\tau)-\mathbf C_x^{\mathrm T}(\tau)\)
		at any \(\tau>0\) (Appendix~\ref{app:finite-lag-asymmetry}).
		This spectral representation further connects the asymmetry of short-time
		cross-correlations to the \(\tau\to0^+\) limit of the violation term in a
		recently derived generalized fluctuation--dissipation
		relation~\cite{Fyodorov2025}
		(Appendix~\ref{app:generalized-fdt}).
		
		Furthermore, heat currents at steady state provide a complementary
		thermodynamic measure of collective nonequilibrium
		dynamics~\cite{Ciliberto2013}.
		With the mobility set to unity, the mean power delivered by the
		deterministic force at node \(m\) is
		\(J_m=-\langle(\mathbf A\mathbf x)_m\circ\dot x_m\rangle_0\),
		where \(\circ\) denotes the Stratonovich product.
		When the driving comes from independent local thermal reservoirs,
		\(J_m\) is the mean heat current from node \(m\) into its
		reservoir~\cite{Sekimoto1998,Seifert2005,Harada2005}. For a general diffusion matrix \(\mathbf D\), we evaluate this average using the Lyapunov equation to obtain
		\begin{equation}
			\mathbf J
			=
			\operatorname{diag}
			\bigl(\mathbf A\mathbf C_0\mathbf A^{\mathrm T}
			-\mathbf A\mathbf D\bigr)
			=
			\operatorname{diag}(\mathbf A\boldsymbol{\Omega}),
			\label{eq:heat-current-asymmetry}
		\end{equation}
		where \(\mathbf J=(J_1,\ldots,J_N)^{\mathrm T}\)
		(see Appendix~\ref{app:heat-currents} for the derivation).
		This relation directly connects the node heat currents
		to the short-time covariance asymmetry.
		Substituting Eq.~\eqref{eq:modal-time-asymmetry} gives
		\begin{equation}
			\mathbf J
			=
			\sum_{i,j=1}^{N}
			\frac{\lambda_i(\lambda_j^*-\lambda_i)}
			{\lambda_i+\lambda_j^*}
			\Gamma_{ij}
			\left(\mathbf U_i\odot\mathbf U_j^*\right),
			\label{eq:heat-spectral}
		\end{equation}
		where \(\odot\) denotes the Hadamard product.
		Heat currents thus admit the similar mode decomposition into contributions
		from fluctuations of individual modes and correlated fluctuations
		between distinct modes.
		Compared with Eq.~\eqref{eq:modal-time-asymmetry},
		the additional factor \(\lambda_i\) weights these contributions
		differently, so that asymmetry of short-time
		cross-correlations and heat currents
		reflect distinct aspects of the same collective fluctuations.
		
		\subsection{Nonequilibrium response operators and spectral bounds}
        Based on the two nonequilibrium observables introduced above,
        we now derive global measures that quantify the system's
        deviation from equilibrium.
        Consider reciprocal networks
        \(\mathbf A=\mathbf A^{\mathrm T}
        =\mathbf U\boldsymbol{\Lambda}\mathbf U^{\mathrm T}\)
        with independent stochastic driving
        \(\mathbf D=\operatorname{diag}(\mathbf d)\),
        where \(\mathbf d=(d_1,\ldots,d_N)^{\mathrm T}\) and \(d_m>0\).
        Uniform driving gives equilibrium, while differences in \(d_m\)
        between coupled nodes provide a nonequilibrium drive,
        analogous to temperature differences between thermal
        reservoirs~\cite{Seifert2012,Godreche2019}.
        To quantify the time-reversal asymmetry induced by heterogeneous
        driving, we define the global asymmetry intensity as
        \(\omega=\|\boldsymbol{\Omega}\|_{\mathrm F}^{2}
        =\sum_{m,n}\Omega_{mn}^{2}\).
        Each entry \(\Omega_{mn}\) measures the asymmetry of short-time
        cross-correlations between nodes \(m\) and \(n\)~\cite{Gnesotto2020,Gradziuk2019,Gradziuk2022}.
        We further quantify thermodynamic dissipation using the entropy
        production rate at steady state \(\sigma=\sum_m J_m/d_m\),
        where \(J_m\) is the mean heat current into the local reservoir
        at node \(m\)~\cite{Seifert2012,Godreche2019}.
        Both global nonequilibrium measures can be expressed in terms of differences
        between local driving intensities,
        \begin{equation}
        	\begin{aligned}
        		\omega
        		&=-\sum_{m<n}\Theta^{\Omega}_{mn}(d_m-d_n)^2,\\
        		\sigma
        		&=-\sum_{m<n}\Theta^{\mathrm J}_{mn}
        		\frac{(d_m-d_n)^2}{d_md_n}.
        	\end{aligned}
        	\label{eq:global-irreversibility}
        \end{equation}
        The matrices \(\boldsymbol{\Theta}^{\Omega}\) and
        \(\boldsymbol{\Theta}^{\mathrm J}\) define two
        \emph{nonequilibrium response operators}
        that connect heterogeneous driving to these nonequilibrium measures.
        For \(\alpha\in\{\Omega,\mathrm J\}\), their off-diagonal entries are
        \begin{equation}
        	\Theta^\alpha_{mn}
        	=-2\int_0^\infty w^\alpha(t)
        	\bigl[(\mathbf A\mathbf G(t))_{mn}\bigr]^2
        	\,\mathrm dt,
        	\label{eq:theta-integral}
        \end{equation}
        where \(w^{\Omega}(t)=2t\), \(w^{\mathrm J}(t)=1\).
        Their equivalent decomposition into mode pairs reads
        \begin{equation}
        	\boldsymbol{\Theta}^{\alpha}
        	=
        	\sum_{i<j}
        	\kappa_{ij}^{\alpha}\boldsymbol{\psi}_{ij},
        	\label{eq:theta-spectral}
        \end{equation}
        where
        \(\boldsymbol{\psi}_{ij}
        =(\mathbf U_i\odot\mathbf U_j)
        (\mathbf U_i\odot\mathbf U_j)^{\mathrm T}\),
        \(\kappa_{ij}^{\Omega}
        =2(\lambda_i-\lambda_j)^2/(\lambda_i+\lambda_j)^2\)
        and
        \(\kappa_{ij}^{\mathrm J}
        =(\lambda_i-\lambda_j)^2/(\lambda_i+\lambda_j)\).
        Both operators are positive semidefinite weighted Laplacians
        determined solely by the network structure \(\mathbf A\)
        (see Appendix~\ref{app:response-operators}). Importantly, the global asymmetry intensity \(\omega\) is an
        exact quadratic form in the driving intensities \(\mathbf d\).
        Near equilibrium or equivalently for weakly heterogeneous driving
        \(\mathbf d=\mathbf1+\delta\mathbf d\) with
        \(\|\delta\mathbf d\|\ll1\), expanding
        Eq.~\eqref{eq:global-irreversibility} about uniform driving
        also gives a quadratic form for the entropy production rate
        to leading order,
        \begin{equation}
        	\begin{aligned}
        		\omega
        		&=\sum_{i<j}\kappa_{ij}^{\Omega}\Gamma_{ij}^{2}
        		=\mathbf d^{\mathrm T}
        		\boldsymbol{\Theta}^{\Omega}\mathbf d,\\
        		\sigma
        		&\simeq\sum_{i<j}\kappa_{ij}^{\mathrm J}\Gamma_{ij}^{2}
        		=\delta\mathbf d^{\mathrm T}
        		\boldsymbol{\Theta}^{\mathrm J}\delta\mathbf d,
        	\end{aligned}
        	\label{eq:quadratic-irreversibility}
        \end{equation}
        where
        \(\Gamma_{ij}
        =(\mathbf U_i\odot\mathbf U_j)^{\mathrm T}\mathbf d
        =(\mathbf U_i\odot\mathbf U_j)^{\mathrm T}\delta\mathbf d\)
        for \(i\neq j\)
        (see Appendix~\ref{app:quadratic-forms}).
        These quadratic forms contain only nonnegative contributions
        from correlated fluctuations between distinct modes
        with different relaxation rates. Using the spectral decomposition
        \begin{equation}
        	\boldsymbol{\Theta}^{\alpha}
        	=\sum_{r=1}^{N}\mu_r^{\alpha}
        	\mathbf v_r^{\alpha}(\mathbf v_r^{\alpha})^{\mathrm T},
        	\label{eq:theta-eigendecomposition}
        \end{equation}
        we obtain orthonormal modes \(\mathbf v_r^{\alpha}\)
        with the ordered spectrum
        \(0=\mu_1^{\alpha}\leq\mu_2^{\alpha}\leq\cdots
        \leq\mu_N^{\alpha}=\mu_{\max}^{\alpha}\).
        The first uniform mode \(\mathbf v_1^{\alpha}=\mathbf1/\sqrt N\)
        corresponds to uniform stochastic driving
        \(\mathbf d\propto\mathbf1\)
        and does not contribute to nonequilibrium.
        The modes with positive eigenvalues define
        activatable modes that can be excited
        by heterogeneous stochastic driving
        \(\delta\mathbf d\perp\mathbf1\), with
        \(\omega=\sum_{r=2}^{N}\mu_r^{\Omega}
        [(\mathbf v_r^{\Omega})^{\mathrm T}\delta\mathbf d]^2\)
        and
        \(\sigma\simeq\sum_{r=2}^{N}\mu_r^{\mathrm J}
        [(\mathbf v_r^{\mathrm J})^{\mathrm T}\delta\mathbf d]^2\).
        Meanwhile, the nonequilibrium response operators determined
        solely by network structure impose fundamental constraints
        on the system's overall deviation from equilibrium. When the uniform mode is the only zero mode of each operator, the Rayleigh quotient yields the following bounds
        for any nonzero \(\delta\mathbf d\),
        \begin{equation}
        	\begin{aligned}
        		\mu_{\min+}^{\Omega}
        		&\leq
        		\frac{\omega}{\delta\mathbf d^{\mathrm T}\delta\mathbf d}
        		\leq
        		\mu_{\max}^{\Omega},\\
        		\mu_{\min+}^{\mathrm J}
        		&\lesssim
        		\frac{\sigma}{\delta\mathbf d^{\mathrm T}\delta\mathbf d}
        		\lesssim
        		\mu_{\max}^{\mathrm J},
        	\end{aligned}
        	\label{eq:theta-eigenspectrum}
        \end{equation}
        where \(\mu_{\min+}^{\alpha}\) and \(\mu_{\max}^{\alpha}\)
        are the smallest positive and largest eigenvalues of
        \(\boldsymbol{\Theta}^{\alpha}\).
        The bounds on \(\omega\) are exact, whereas those on
        \(\sigma\) hold approximately near equilibrium.
        The corresponding extremal stochastic driving patterns are
        \(\delta\mathbf d\propto\mathbf v_{\min+}^{\alpha}\) and
        \(\delta\mathbf d\propto\mathbf v_{\max}^{\alpha}\)
        if the resulting driving intensities satisfy \(d_m>0\) at every node.
        Near equilibrium, the two nonequilibrium response operators
        also constrain the global asymmetry intensity per unit
        entropy production rate,
        \begin{equation}
        	\rho_{\min}
        	\lesssim
        	\frac{\omega}{\sigma}
        	\lesssim
        	\rho_{\max},
        	\label{eq:asymmetry-dissipation-bounds}
        \end{equation}
        where \(\rho_{\min}\) and \(\rho_{\max}\) are the smallest
        and largest generalized eigenvalues of
        \(\boldsymbol{\Theta}^{\Omega}\boldsymbol{\ell}_r
        =\rho_r\boldsymbol{\Theta}^{\mathrm J}\boldsymbol{\ell}_r\)
        on the subspace \(\boldsymbol{\ell}_r\perp\mathbf1\).
        These bounds hold approximately near equilibrium,
        and the corresponding extremal stochastic driving patterns are
        \(\delta\mathbf d\propto\boldsymbol{\ell}_{\min}\)
        and
        \(\delta\mathbf d\propto\boldsymbol{\ell}_{\max}\)
        (see Appendix~\ref{app:generalized-eigenvalue-bounds}).
        Numerical validation of the bounds in
        Eqs.~\eqref{eq:theta-eigenspectrum}
        and~\eqref{eq:asymmetry-dissipation-bounds}
        is shown in Fig.~S1 of the SM~\cite{SupplementalMaterial}.
        
\begin{figure}[t]
	    	\centering
	    	\includegraphics[width=\columnwidth]{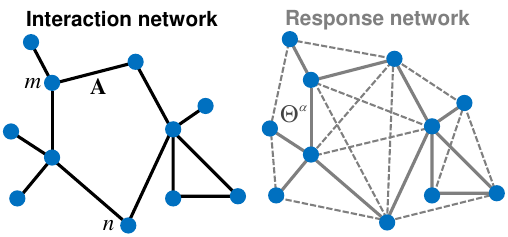}
	    	\caption{
	    		Network representation of the nonequilibrium response operators.
	    		Black solid lines (left) represent direct interactions
	    		described by \(\mathbf A\).
	    		Gray lines (right) represent selected nonzero off-diagonal
	    		entries of \(\boldsymbol{\Theta}^{\alpha}\),
	    		with \(\alpha\in\{\Omega,\mathrm J\}\).
	    		Gray solid lines connect directly interacting nodes,
	    		while gray dashed lines connect nodes without a direct
	    		interaction, such as \(m\) and \(n\).
	    		The response operators depend only on \(\mathbf A\),
	    		independently of stochastic driving, and incorporate
	    		propagation through both direct and indirect interactions.
	    		Their associated response networks are therefore generally
	    		dense even when the interaction network is sparse.
	    	}
	    	\label{fig:structure-response}
	    \end{figure}

        Together, these results show how network topology constrains the
        range of collective nonequilibrium responses to stochastic driving.
        The mapping
        \(\mathbf A\mapsto
        \{\boldsymbol{\Theta}^{\mathrm J},\boldsymbol{\Theta}^{\Omega}\}\)
        separates structural effects from the imposed driving.
        Through the network propagator, these nonequilibrium response operators incorporate
        dynamical propagation between nodes through direct and
        indirect interactions and are generally dense even when
        the underlying network is sparse
        [Fig.~\ref{fig:structure-response}].
        Their eigenvectors \(\mathbf v_r^\alpha\) identify stochastic
        driving patterns. At fixed heterogeneous driving strength,
        their spectra \(\{\mu_r^\alpha\}\) determine the attainable
        upper and lower bounds on the global asymmetry intensity and
        on the entropy production rate near equilibrium.
        The bounds are exact for \(\omega\) and approximate
        near equilibrium for \(\sigma\).
        The generalized spectrum \(\{\rho_r\}\) further determines
        the approximate range of \(\omega/\sigma\) near equilibrium,
        with the corresponding extremal driving patterns given by
        \(\boldsymbol{\ell}_{\min}\) and \(\boldsymbol{\ell}_{\max}\).
 
	    \section{Hierarchical modular networks and the human brain}
	    
\begin{figure*}[t]
	    	\centering
	    	\includegraphics{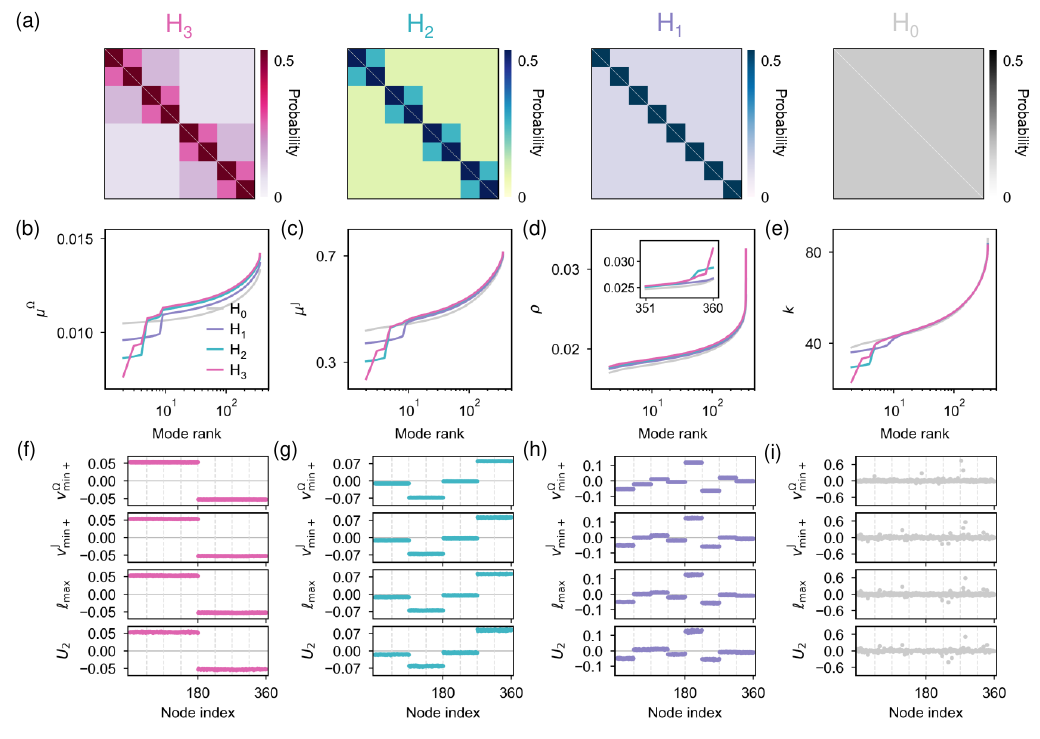}
	    	\caption{
	    		Spectra of nonequilibrium response operators and
	    		extremal stochastic driving patterns in synthetic
	    		hierarchical modular networks.
	    		All realizations are binary and undirected.
	    		The numbers of nodes and connections are fixed
	    		across \(\mathrm{H}_3\)--\(\mathrm{H}_0\) (see SM for network generation), with
	    		\(N=360\) and mean degree \(\langle e\rangle=60\),
	    		ensuring the same wiring cost.
	    		The dynamical parameters are \(\gamma=g=1\).
	    		(a) Matrices of connection probabilities between
	    		node pairs used to generate the adjacency matrices
	    		\(\mathbf W_{\mathrm{H}_3}\)--\(\mathbf W_{\mathrm{H}_0}\).
	    		\(\mathrm{H}_3\) contains three nested levels with
	    		\(8\), \(4\), and \(2\) modules.
	    		\(\mathrm{H}_2\) and \(\mathrm{H}_1\) retain the
	    		finer two and one levels, respectively, while
	    		\(\mathrm{H}_0\) is a homogeneous random network.
	    		(b--e) Spectra \(\{\mu_r^\Omega\}\),
	    		\(\{\mu_r^{\mathrm{J}}\}\), \(\{\rho_r\}\),
	    		and \(\{k_i\}\), respectively.
	    		Eigenvalues are sorted within each realization
	    		and averaged at each rank over \(100\) realizations.
	    		Uniform zero modes are omitted from the response
	    		operator and Laplacian spectra.
	    		The inset in (d) enlarges the spectra near
	    		the upper generalized bound.
	    		(f--i) Extremal stochastic driving patterns for
	    		\(\mathrm{H}_3\)--\(\mathrm{H}_0\), showing
	    		\(\mathbf v_{\min+}^{\Omega}\),
	    		\(\mathbf v_{\min+}^{\mathrm{J}}\),
	    		\(\boldsymbol{\ell}_{\max}\), and the Laplacian
	    		Fiedler mode \(\mathbf U_2\) from top to bottom.
	    		For each network class, the adjacency matrices
	    		of the same \(100\) realizations are first averaged.
	    		The extremal driving patterns and Fiedler mode
	    		are then computed for the resulting mean network.
	    		Dashed vertical lines delimit eight reference
	    		groups of \(45\) nodes.
	    		Similar trends in the spectral bounds are observed
	    		under variations in mean degree, network size,
	    		and dynamical parameters \(\gamma\) and \(g\)
	    		(SM, Figs.~S2--S5).
	    	}
	    	\label{fig:hierarchical-networks}
	    \end{figure*}

	    We can now use the two nonequilibrium response operators to compare
	    how different network topologies influence collective nonequilibrium
	    dynamics, focusing on their spectra \(\{\mu_r^\alpha\}\),
	    the generalized spectrum \(\{\rho_r\}\) and the associated
	    extremal stochastic driving patterns.
	    Of particular interest are hierarchical modular networks,
	    which are widespread in nature and include human structural
	    connectomes~\cite{Bassett2010,Wang2019}.
	    In these networks, strongly interconnected groups of nodes form
	    weakly coupled modules that are nested within progressively
	    larger modules across multiple scales
	    [Fig.~\ref{fig:hierarchical-networks}(a)]. Such topologies have been shown to confer various functional
	    advantages~\cite{Bassett2010,Wang2019}, yet their potential benefits
	    for nonequilibrium operation remain unclear.
	    More importantly, previous studies have shown that resting-state
	    brain activity is maintained in specific nonequilibrium
	    states~\cite{Gilson2023,Deco2023}.
	    Since brain activity is thought to be strongly constrained by
	    the underlying structural connectivity network~\cite{Wang2019}, these findings
	    raise the question of which intrinsic modes are excited to
	    sustain these nonequilibrium states.
	    To address these two questions, we consider diffusive dynamics
	    on an undirected network with nonnegative adjacency matrix
	    \(\mathbf W\), described by
	    \(\mathbf A=\gamma\mathbf I+g\mathbf L\)~\cite{Korniss2007,Ren2010,Ronellenfitsch2018}.
	    Here, \(\mathbf L=\operatorname{diag}(\mathbf W\mathbf1)-\mathbf W\)
	    is the graph Laplacian, while \(\gamma\) and \(g\) denote the
	    local relaxation rate and global coupling strength, respectively.
	    The dynamics are stable for \(\gamma,g>0\).
	    To compare different networks, we hold \(\gamma\) and \(g\)
	    fixed and match their wiring costs. For resting-state brain activity, we infer the stochastic driving
	    patterns and investigate which intrinsic modes are activated
	    to sustain these specific nonequilibrium states.
	    
	    \subsection{Synthetic hierarchical modular networks}
	    
	    We first construct binary synthetic hierarchical modular
	    networks ranging from \(\mathrm{H}_3\) with three nested
	    modular levels to the homogeneous random network
	    \(\mathrm{H}_0\)
	    [Fig.~\ref{fig:hierarchical-networks}(a)].
	    As hierarchical levels are removed from
	    \(\mathrm{H}_3\) to \(\mathrm{H}_0\), both spectra
	    \(\{\mu_r^\alpha\}\) become more concentrated as their
	    lower bounds \(\mu_{\min+}^{\Omega}\) and
	    \(\mu_{\min+}^{\mathrm{J}}\) increase, while their
	    upper bounds remain comparatively unchanged
	    [Fig.~\ref{fig:hierarchical-networks}(b,c)].
	    Meanwhile, the upper generalized bound
	    \(\rho_{\max}\) decreases
	    [Fig.~\ref{fig:hierarchical-networks}(d)].
	    For these diffusive dynamics, the relaxation modes
	    are the orthonormal Laplacian modes satisfying
	    \(\mathbf L\mathbf U_i=k_i\mathbf U_i\), with
	    \(0=k_1<k_2\leq\cdots\leq k_N\) and
	    \(\lambda_i=\gamma+gk_i\).
	    The lower Laplacian bound \(k_2\) also increases
	    as hierarchical levels are removed
	    [Fig.~\ref{fig:hierarchical-networks}(e)].
	    For the mean networks of classes
	    \(\mathrm{H}_3\)--\(\mathrm{H}_1\), the extremal
	    stochastic driving patterns reflect the modular
	    partitions
	    [Fig.~\ref{fig:hierarchical-networks}(f--h)].
	    Specifically, \(\mathbf v_{\min+}^{\Omega}\),
	    \(\mathbf v_{\min+}^{\mathrm{J}}\), and
	    \(\boldsymbol{\ell}_{\max}\) closely align with
	    the \emph{Fiedler mode} \(\mathbf U_2\), associated
	    with the algebraic connectivity \(k_2\).
	    This mode describes the slowest nonuniform relaxation
	    and is used in spectral partitioning to identify
	    weakly coupled modules~\cite{Fiedler1973}.
	    The random \(\mathrm{H}_0\) network shows no comparable
	    common alignment of these driving patterns with
	    the Fiedler mode
	    [Fig.~\ref{fig:hierarchical-networks}(i)].
	    
	    In the ideal hierarchical networks, this alignment
	    becomes exact, with the two lower response bounds
	    and the upper generalized bound attained by a common
	    driving direction
	    \(\delta\mathbf d
	    \parallel\mathbf v_{\min+}^{\Omega}
	    \parallel\mathbf v_{\min+}^{\mathrm{J}}
	    \parallel\boldsymbol{\ell}_{\max}
	    \parallel\mathbf U_2\).
	    For example, ideal \(\mathrm{H}_3\) has uniform
	    coupling between its two equally sized largest modules.
	    Driving along \(\mathbf U_2\) imposes different
	    intensities between these modules while keeping
	    the intensity uniform within each.
	    The resulting nonequilibrium response receives
	    contributions only from correlated fluctuations
	    between the uniform mode
	    \(\mathbf U_1=\mathbf1/\sqrt N\) and the Fiedler
	    mode \(\mathbf U_2\).
	    The corresponding spectral bounds are
	    \(\mu_{\min+}^{\Omega}
	    =2(gk_2)^2/[N(2\gamma+gk_2)^2]\),
	    \(\mu_{\min+}^{\mathrm{J}}
	    =(gk_2)^2/[N(2\gamma+gk_2)]\), and
	    \(\rho_{\max}=2/(2\gamma+gk_2)\)
	    (see Appendix~\ref{app:fiedler-extrema}).
	    Increasing \(k_2\) therefore raises both lower
	    response bounds and reduces the upper generalized
	    bound, consistent with the spectral changes observed
	    as hierarchical levels are removed.

	    \begin{figure*}[t]
	    	\centering
	    	\includegraphics{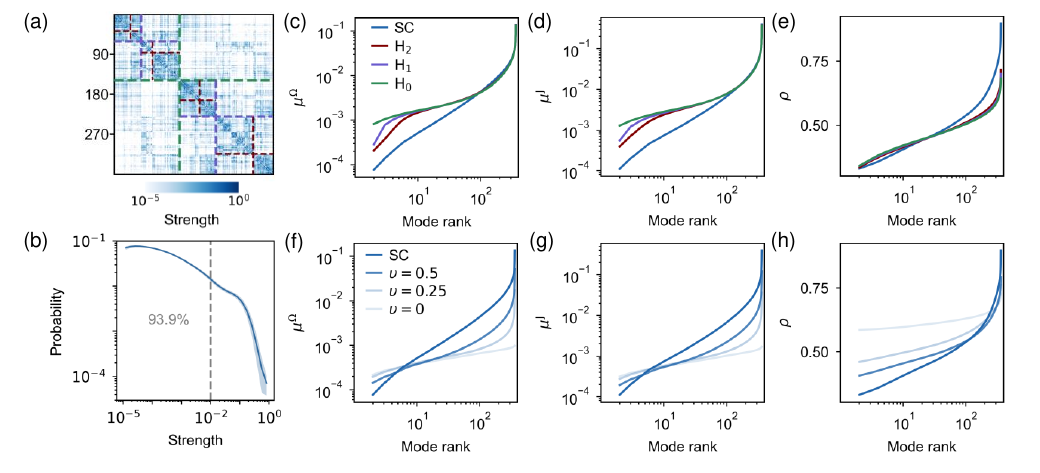}
	    	\caption{
	    		Spectra of nonequilibrium response operators in human
	    		structural connectomes and their surrogate networks.
	    		(a) An individual SC network reordered according to nested
	    		spectral partitions~\cite{Wang2019}.
	    		Dashed lines mark module boundaries.
	    		Weights are normalized by the maximum connection weight
	    		for each participant.
	    		(b) Mean distribution of nonzero connection weights,
	    		with shading indicating one standard deviation
	    		across participants.
	    		The dashed line marks \(10^{-2}\), below which
	    		\(93.9\%\) of connections lie on average.
	    		(c--e) Spectra \(\{\mu_r^\Omega\}\),
	    		\(\{\mu_r^{\mathrm{J}}\}\), and \(\{\rho_r\}\),
	    		respectively, for SC and hierarchical modular
	    		randomization (HMR) surrogates.
	    		HMR randomly redistributes existing weighted
	    		connections within the finer modules
	    		(\(\mathrm{H}_2\)), the two largest modules
	    		(\(\mathrm{H}_1\)), or the whole network
	    		(\(\mathrm{H}_0\)), preserving connections
	    		between selected modules.
	    		(f--h) Corresponding spectra for hierarchical
	    		modular homogenization (HMH) surrogates.
	    		HMH adjusts nonzero weights toward their mean
	    		across the whole network while preserving
	    		connection positions.
	    		Progressively lighter blue curves denote
	    		\(\upsilon=1\) (SC), \(0.5\), \(0.25\), and \(0\),
	    		with \(\upsilon=0\) assigning uniform weights
	    		to all existing connections.
	    		Each surrogate preserves the connection count
	    		and total weight of its empirical counterpart.
	    		All networks contain \(N=360\) cortical regions,
	    		with diffusive dynamics
	    		\(\mathbf A=\gamma\mathbf I+g\mathbf L\)
	    		and \(\gamma=g=1\).
	    		Eigenvalues are sorted within each participant
	    		and averaged at each rank over \(993\) participants.
	    	}
	    	\label{fig:brain-spectrum}
	    \end{figure*}
	    
	    \subsection{Human brain structural connectomes}
	    
	    We next examine how the topology of human structural
	    connectomes shapes the spectra of the two nonequilibrium
	    response operators.
	    Using DWI data from \(993\) participants in the
	    Human Connectome Project~\cite{VanEssen2013}, we construct individual weighted
	    and undirected structural connectivity (SC) networks of
	    \(360\) cortical regions and obtain \(360\times360\)
	    adjacency matrices (see SM for data processing).
	    The SC networks exhibit modular organization with
	    multiple nested levels
	    [Fig.~\ref{fig:brain-spectrum}(a)].
	    More importantly, SC networks exhibit multiscale
	    connection weights, with predominantly weak connections
	    (e.g., \(93.9\%\) have weights below \(10^{-2}\) on average)
	    and a small fraction of strong connections
	    [Fig.~\ref{fig:brain-spectrum}(b)].
	    To examine whether SC networks also exhibit broad
	    spectral bounds, we use them to construct two families
	    of surrogate networks.
	    (i) In hierarchical modular randomization (HMR),
	    the existing weighted connections within each module
	    at hierarchical level \(\mathrm{H}_n\) are redistributed
	    uniformly at random among all possible node pairs
	    within that module, while connections between modules
	    remain unchanged. This disrupts the finer modular organization at level
	    \(\mathrm{H}_{n+1}\) while preserving the connection count
	    and total connection weight within each \(\mathrm{H}_n\) module.
	    (ii) In hierarchical modular homogenization (HMH),
	    nonzero weights are transformed as
	    \(W_{ij}^{(\upsilon)}=\upsilon W_{ij}+(1-\upsilon)W_m\),
	    where \(W_m\) is the mean weight of existing connections
	    and \(0\leq\upsilon\leq1\).
	    Here, \(\upsilon=1\) recovers SC and \(\upsilon=0\)
	    gives uniform weights across all existing connections,
	    while preserving connection positions, count and total weight.
	    Therefore, HMR disrupts the hierarchical modular
	    organization of SC, whereas HMH homogenizes its
	    multiscale connection weights
	    (see SM for surrogate network construction).
	    
	    In HMR, both spectra
	    \(\{\mu_r^\alpha\}\) become more concentrated as their
	    lower bounds \(\mu_{\min+}^{\Omega}\) and
	    \(\mu_{\min+}^{\mathrm J}\) increase, while their
	    upper bounds remain comparatively unchanged
	    [Fig.~\ref{fig:brain-spectrum}(c,d)].
	    The generalized spectrum also narrows, primarily through
	    a reduction in \(\rho_{\max}\), with comparatively little
	    change in \(\rho_{\min}\)
	    [Fig.~\ref{fig:brain-spectrum}(e)]. 
	    These changes parallel those observed when hierarchical
	    levels are removed from the synthetic networks, with
	    an accompanying increase in \(k_2\) (Fig.~S7).
	    In HMH, both spectra
	    \(\{\mu_r^\alpha\}\) become more concentrated as their
	    upper bounds \(\mu_{\max}^{\Omega}\) and
	    \(\mu_{\max}^{\mathrm J}\) decrease, while
	    changes in their lower bounds are comparatively smaller
	    [Fig.~\ref{fig:brain-spectrum}(f,g)]. The generalized spectrum contracts from both ends,
	    with pronounced upward and downward shifts in
	    \(\rho_{\min}\) and \(\rho_{\max}\), respectively
	    [Fig.~\ref{fig:brain-spectrum}(h)].
	    Complete weight homogenization (\(\upsilon=0\))
	    produces the narrowest spectral ranges.
	    These trends are also reproduced in synthetic weighted
	    hierarchical modular networks as connection weights
	    become more uniform (Fig.~S6).
	    
	    Together, the results for synthetic networks and human structural
	    connectomes show that hierarchical modular organization and
	    multiscale connection weights jointly broaden the nonequilibrium
	    operating range at fixed wiring cost.
	    Hierarchy extends both response spectra toward smaller nonzero
	    eigenvalues, whereas multiscale connection weights sustain larger
	    upper spectral bounds.
	    At fixed driving heterogeneity, these complementary effects
	    permit both lower and higher global asymmetry intensities,
	    as well as a broader range of entropy production rates near
	    equilibrium.
	    Hierarchy also raises the upper bound of the generalized spectrum,
	    while multiscale connection weights reduce its lower bound,
	    expanding the attainable range of global asymmetry intensity
	    per unit entropy production rate near equilibrium.
	    
	    \begin{figure*}[t]
	    	\centering
	    	\includegraphics{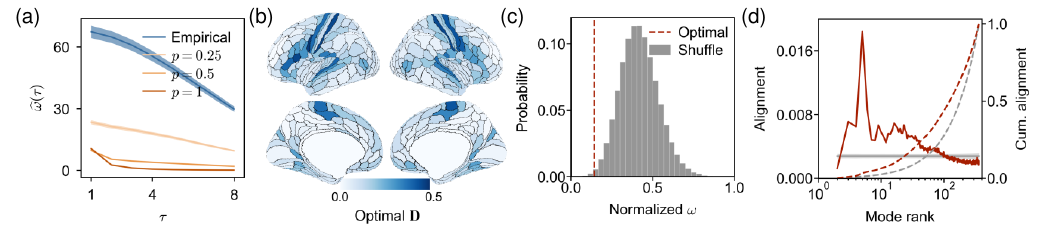}
	    	\caption{
	    		Nonequilibrium dynamics and inferred stochastic driving
	    		in the resting human brain.
	    		(a) Global asymmetry intensity \(\widehat{\omega}(\tau)\)
	    		estimated from empirical resting-state fMRI time series
	    		(blue) and temporally shuffled series with \(p=0.25\),
	    		\(0.5\), and \(1\) (orange shades).
	    		A randomly selected fraction \(p\) of time points is
	    		shuffled jointly across all brain regions.
	    		Lines show participant means, with shading indicating
	    		pointwise \(95\%\) confidence intervals.
	    		(b) Group average of the optimal stochastic driving patterns inferred
	    		by fitting the model to each participant's resting-state fMRI data.
	    		Regional driving intensities \(d_m\) are min--max normalized to \([0,1]\)
	    		within each participant before averaging across participants.
	    		(c) Global asymmetry intensity \(\omega\) for optimal
	    		and spatially shuffled driving patterns.
	    		The operator \(\boldsymbol{\Theta}^{\Omega}\) is
	    		recalculated from each participant's structural connectome
	    		using the optimal \(\gamma\) and \(g\).
	    		Spatial shuffling reassigns regional driving intensities
	    		while preserving their distribution and keeping the
	    		operator fixed.
	    		The optimal and \(10^5\) shuffled values of \(\omega\)
	    		are jointly min--max normalized to \([0,1]\) within
	    		each participant.
	    		The gray histogram pools normalized shuffled values
	    		across participants, and the red dashed line marks
	    		the mean normalized intensity of the optimal patterns. We use a one-sided permutation test to assess whether the mean
	    		normalized \(\omega\) across participants is lower for the optimal
	    		patterns, computing one group mean per shuffle.
	    		(d) Alignment strengths \(\chi_r\) of optimal (red) and spatially
	    		shuffled (gray) driving patterns with the nonuniform modes of
	    		\(\boldsymbol{\Theta}^{\Omega}\).
	    		Nonuniform modes are ordered by increasing eigenvalue
	    		within each participant.
	    		Solid and dashed curves show mean alignment strength
	    		(left axis) and mean cumulative alignment strength
	    		(right axis), respectively.
	    		Gray shading indicates pointwise central \(95\%\)
	    		intervals of the group shuffle distributions.
	    		Results are based on \(993\) participants and \(N=360\) brain regions. The Supplemental Material~\cite{SupplementalMaterial} describes
	    		MRI acquisition and processing, cortical parcellation, and
	    		structural connectome reconstruction from diffusion MRI
	    		(see Refs.~\cite{Smith2013HCP,Glasser2013,Glasser2016,Tournier2019, Dhollander2016,
	    			Dhollander2019,Jeurissen2014,Smith2012ACT,
	    			Tournier2010,Smith2015SIFT2} for the methods used).
	    	}
	    	\label{fig:brain-resting}
	    \end{figure*}
	    
	    \subsection{Nonequilibrium dynamics in the resting-state human brain}
	    
	    To investigate whether resting-state brain activity is out of
	    equilibrium, we first estimate the global asymmetry intensity
	    \(\widehat{\omega}(\tau)\) at multiple time lags from each
	    participant's fMRI time series
	    (Appendix~\ref{app:driving-inference}).
	    The empirical intensity decreases with lag but remains higher
	    than that of temporally shuffled time series throughout the
	    examined range.
	    These shuffled series are generated by shuffling a randomly
	    selected fraction \(p\) of the time points, using the same
	    temporal shuffle across all brain regions.
	    Increasing \(p\) from \(0.25\) to \(0.5\) and \(1\) progressively
	    reduces the intensity [Fig.~\ref{fig:brain-resting}(a)],
	    indicating that spontaneous resting-state brain activity
	    sustains temporally organized nonequilibrium dynamics.
	    
	    We next use the model to infer the stochastic driving pattern
	    associated with resting-state brain activity and examine its role
	    in sustaining these nonequilibrium dynamics.
	    Under the assumption of independent stochastic driving across
	    brain regions with \(\mathbf D=\operatorname{diag}(\mathbf d)\),
	    the optimal stochastic driving pattern \(\mathbf d\) and parameters
	    \((\gamma,g)\) are obtained by minimizing the joint fitting error
	    between the model and empirical zero-lag covariances and one-lag
	    covariance asymmetries (Appendix~\ref{app:driving-inference}).
	    The optimal patterns are spatially heterogeneous, as illustrated
	    by their group-averaged distribution
	    [Fig.~\ref{fig:brain-resting}(b)].
	    Optimal parameter estimates and fitting errors are reported
	    in Fig.~S8.
	    Similar spatial driving patterns are obtained from fits
	    including additional lags (Fig.~S9).
	    To determine how this spatial organization shapes the global
	    asymmetry intensity, we recalculate the nonequilibrium response
	    operator \(\boldsymbol{\Theta}^{\Omega}\) and its spectrum
	    from each participant's structural connectome using the optimal
	    \(\gamma\) and \(g\).
	    Keeping the operator fixed, we compare
	    \(\omega=\mathbf d^{\mathrm T}\boldsymbol{\Theta}^{\Omega}\mathbf d\)
	    for each participant's optimal pattern with the values obtained
	    by spatially shuffling its driving intensities across brain regions
	    \(10^5\) times.
	    The optimal patterns yield a lower mean \(\omega\) across participants
	    than the shuffled patterns (permutation test, \(p<10^{-4}\))
	    [Fig.~\ref{fig:brain-resting}(c)]. The entropy production rate \(\sigma\) is likewise lower for the
	    optimal patterns (Fig.~S10).
	    To explain this difference, we examine whether the optimal
	    patterns preferentially align with intrinsic modes associated
	    with smaller eigenvalues of \(\boldsymbol{\Theta}^{\Omega}\).
	    Since the uniform component does not contribute to \(\omega\),
	    the optimal pattern is centered as
	    \(\delta\mathbf d=\mathbf d-\bar d\,\mathbf1\) for each participant.
	    We order the nonuniform modes \(r=2,\ldots,N\) by increasing
	    eigenvalue and define the alignment strength with mode \(r\)
	    as the normalized squared projection
	    \(\chi_r=[(\mathbf v_r^{\Omega})^{\mathrm T}\delta\mathbf d]^2
	    /\|\delta\mathbf d\|_2^2\), with \(\sum_{r=2}^{N}\chi_r=1\).
	    The exact relation
	    \(\omega/\|\delta\mathbf d\|_2^2
	    =\sum_{r=2}^{N}\mu_r^{\Omega}\chi_r\)
	    shows that, at fixed driving heterogeneity, shifting alignment
	    from modes with larger eigenvalues to those with smaller
	    eigenvalues lowers the global asymmetry intensity.
	    We found that the optimal patterns exhibit greater alignment
	    strengths at the lower end of the spectrum, whereas spatially
	    shuffled patterns show lower values at these mode ranks and
	    approximately uniform mean alignment across the spectrum.
	    The cumulative alignment strength of the optimal patterns
	    also rises more rapidly at low mode ranks
	    [Fig.~\ref{fig:brain-resting}(d)].
	    Fits including additional lags show consistent results in both
	    the spatial shuffle and alignment analyses (Fig.~S11).
	    This preferential alignment accounts for their lower global
	    asymmetry intensity.
	    Thus, such spatial organization of stochastic driving may
	    help maintain resting-state brain activity at a comparatively
	    low nonequilibrium level.
	    
		\section{Discussion}
		Stochastic thermodynamics has made substantial progress in
		addressing inverse problems, enabling the inference of underlying
		dynamical mechanisms and thermodynamic properties of nonequilibrium
		systems from observed fluctuations.
		Specifically, previous studies have inferred dissipation bounds
		from current fluctuations~\cite{Gingrich2017}, constrained
		thermodynamic affinities through cross-correlation
		asymmetry~\cite{Ohga2023} and reconstructed force fields and
		noise intensities from observed
		dynamics~\cite{Frishman2020,Gnesotto2020}.
		By contrast, although previous studies have examined the effects
		of network topology and stochastic driving on nonequilibrium
		systems~\cite{NartalloKaluarachchi2024,Mura2018,Gradziuk2019},
		the forward problem of determining how network topology itself
		shapes collective responses to different stochastic driving
		patterns remains insufficiently explored.
		Here, starting from the principle of microscopic reversibility,
		we derive a mode decomposition that identifies the contributions
		of fluctuations of individual modes and correlated fluctuations
		between distinct modes to the asymmetry of short-time
		cross-correlations and heat currents at steady state.
		This connects mode correlations in probe filaments embedded
		in active networks~\cite{Gladrow2017} with recent results on
		nonorthogonal modes and entropy production~\cite{Fyodorov2025}.
		A complementary analysis of Ornstein--Uhlenbeck processes
		separates the entropy production rate at steady state into
		oscillatory and nonnormal contributions~\cite{Nagayama2026}.
		For reciprocal linear networks, the nonequilibrium response
		operators characterize how network topology constrains collective
		nonequilibrium responses across stochastic driving patterns. The generalized spectrum further relates the global asymmetry
		intensity to the entropy production rate near equilibrium.
		Therefore, our study may help guide the design and tuning of network
		topology to achieve targeted operating points and ranges in
		diverse nonequilibrium systems~\cite{Gladrow2017,Ciliberto2013}.
		
		The human brain connectome exhibits hierarchical modular
		organization, which has been shown to confer several functional
		advantages, including dynamical stability under connectivity
		changes~\cite{Robinson2009}, enhanced sensitivity to
		stimuli~\cite{Moretti2013} and greater functional
		diversity~\cite{Wang2019}.
		Here, we show that these advantages extend to nonequilibrium
		dynamics, as hierarchical modular organization broadens the
		response spectra and expands the nonequilibrium operating
		range under stochastic driving.
		This broader range may help accommodate the distinct
		nonequilibrium regimes observed during resting-state and
		task-related brain activity~\cite{Kringelbach2023,Deco2023}.
		Furthermore, our findings suggest a possible mechanism for
		the low nonequilibrium level reported at rest relative to
		cognitive tasks.
		Studies of connectome harmonics link spontaneous activity to
		anatomically constrained spatial patterns~\cite{Atasoy2016},
		while task studies show that functional networks largely retain
		the intrinsic organization present at rest~\cite{Cole2014}.
		Our results extend this connection between structure and activity
		to the spatial organization of stochastic driving.
		The inferred resting-state driving patterns preferentially align
		with modes of \(\boldsymbol{\Theta}^{\Omega}\) associated with
		smaller eigenvalues. Compared with spatial shuffles of the same
		regional driving intensities, the inferred patterns yield lower
		global asymmetry intensity and a lower entropy production rate.
		This alignment suggests a possible mechanism for the low
		nonequilibrium level reported at rest relative to cognitive
		tasks~\cite{Kringelbach2023,Deco2023}.
		Redistributing regional driving could recruit different
		intrinsic modes and adjust the nonequilibrium level, offering
		a possible dynamical basis for the functional reconfiguration
		associated with changing cognitive
		demands~\cite{Cole2014,Shine2016}.
		
		In summary, network topology constrains the range of collective
		nonequilibrium responses at fixed driving heterogeneity, while
		the spatial pattern of stochastic driving selects an operating
		point within that range.
		Importantly, the nonequilibrium response operators could support
		broader investigations of reciprocal linear networks with
		hierarchical modular, small-world~\cite{Watts1998} and
		scale-free organization~\cite{Barabasi1999}.
		They could be used to identify structural features that expand
		or restrict the nonequilibrium operating range, assess how
		changes in connectivity alter its limits, and predict how
		redistributing local driving shifts a system's operating point.
		Extending these operators to temporally correlated
		driving~\cite{Gradziuk2022} would further allow the spatial
		and temporal organization of stochastic inputs to be studied
		together, revealing how driving correlation times interact
		with network relaxation to favor different contributions
		from intrinsic modes.
		Such developments would provide testable predictions for how
		network structure and stochastic inputs jointly determine
		the nonequilibrium operating states of biological and
		engineered systems.

        \begin{acknowledgments}
        	We thank Professor Changsong Zhou for inspiring this work
        	and for insightful discussions.
        	This work was supported by the National Natural Science Foundation
        	of China under Grant Nos.~12132012, 32541016, and 12272292.
        \end{acknowledgments}

       \section*{Data Availability}
       The neuroimaging data are available from the Human Connectome
       Project Young Adult dataset~\cite{HCPYoungAdultData}.
       The data generated in this study are available from the authors
       upon reasonable request.

        \appendix
        \section{Mode decomposition of the asymmetry of cross-correlations
        	at arbitrary time lags}
        \label{app:finite-lag-asymmetry}
        
        At steady state, the zero-lag covariance matrix in mode coordinates
        satisfies
        \(\boldsymbol{\Lambda}\mathbf C_y(0)
        +\mathbf C_y(0)\boldsymbol{\Lambda}^{\dagger}
        =2\boldsymbol{\Gamma}\).
        Combining this relation with the lagged propagation
        \(\mathbf C_y(\tau)=e^{-\boldsymbol{\Lambda}\tau}\mathbf C_y(0)\)
        gives the cross-correlations between modes,
        \begin{equation}
        	C_{y_i y_j}(\tau)
        	=
        	\frac{2\Gamma_{ij}}{\lambda_i+\lambda_j^*}
        	e^{-\lambda_i\tau}.
        	\label{eq:appendix-mode-covariance}
        \end{equation}
        Transforming back to node coordinates through
        \(\mathbf C_x(\tau)=\mathbf U\mathbf C_y(\tau)\mathbf U^\dagger\)
        yields the asymmetry of lagged cross-correlations,
        \begin{equation}
        	\mathbf C_x(\tau)-\mathbf C_x^{\mathrm T}(\tau)
        	=2\sum_{i,j=1}^{N}
        	\frac{e^{-\lambda_i\tau}-e^{-\lambda_j^*\tau}}
        	{\lambda_i+\lambda_j^*}
        	\Gamma_{ij}\mathbf U_i\mathbf U_j^\dagger.
        	\label{eq:finite-lag-asymmetry}
        \end{equation}
        This decomposition is exact for any \(\tau>0\).
        The terms with \(i=j\) describe contributions from fluctuations
        of individual modes, whereas those with \(i\neq j\) describe
        contributions from correlated fluctuations between distinct modes.
        
        To connect this finite-lag expression to the short-time
        asymmetry \(\boldsymbol{\Omega}\), we use
        \begin{equation}
        	\lim_{\tau\to0^+}
        	\frac{e^{-\lambda_i\tau}-e^{-\lambda_j^*\tau}}{\tau}
        	=\lambda_j^*-\lambda_i.
        	\label{eq:short-time-exponential-limit}
        \end{equation}
        Dividing Eq.~\eqref{eq:finite-lag-asymmetry} by \(2\tau\)
        and taking \(\tau\to0^+\) therefore recovers the mode
        decomposition of \(\boldsymbol{\Omega}\) in
        Eq.~\eqref{eq:modal-time-asymmetry}.
        
        \section{Asymmetry of short-time cross-correlations and the
        	generalized fluctuation--dissipation relation}
        \label{app:generalized-fdt}
        
        To relate \(\boldsymbol{\Omega}\) to the generalized
        fluctuation--dissipation relation of Ref.~\cite{Fyodorov2025},
        we introduce a weak external force \(\mathbf f(t)\) into the
        Langevin dynamics,
        \begin{equation}
        	\dot{\mathbf x}(t)
        	=
        	-\mathbf A\mathbf x(t)
        	+\boldsymbol{\eta}(t)
        	+\mathbf f(t).
        	\label{eq:si-fdr-perturbed}
        \end{equation}
        The resulting change in the mean is
        \begin{equation}
        	\delta\langle\mathbf x(t)\rangle
        	=
        	\int_{-\infty}^{t}
        	\mathbf R(t-s)\mathbf f(s)\,\mathrm ds,
        	\label{eq:si-fdr-response}
        \end{equation}
        where \(\mathbf R(\tau)=e^{-\mathbf A\tau}H(\tau)\)
        is the causal response matrix and \(H(\tau)\) is the
        Heaviside function.
        In the unperturbed steady state, the lagged cross-correlation
        matrix satisfies
        \(\mathbf C_x(\tau)=e^{-\mathbf A\tau}\mathbf C_0\)
        for \(\tau>0\), with \(\mathbf C_0=\mathbf C_x(0)\).
        Differentiating this expression and multiplying the Lyapunov
        equation, Eq.~\eqref{eq:lyapunov}, from the left by
        \(\mathbf R(\tau)\) give
        \begin{equation}
        	\begin{aligned}
        		\partial_\tau\mathbf C_x(\tau)
        		&=-\mathbf A\mathbf C_x(\tau),\\
        		2\mathbf R(\tau)\mathbf D
        		&=\mathbf A\mathbf C_x(\tau)
        		+\mathbf C_x(\tau)\mathbf A^{\mathrm T},
        	\end{aligned}
        	\label{eq:si-fdr-identities}
        \end{equation}
        where the second identity uses the commutation of
        \(\mathbf A\) with its propagator.
        Combining these identities yields the generalized
        fluctuation--dissipation relation,
        \begin{equation}
        	2\partial_\tau\mathbf C_x(\tau)
        	=
        	-2\mathbf R(\tau)\mathbf D
        	+\boldsymbol{\Delta}(\tau),
        	\label{eq:si-fdr-generalized}
        \end{equation}
        where
        \begin{equation}
        	\boldsymbol{\Delta}(\tau)
        	=
        	\mathbf C_x(\tau)\mathbf A^{\mathrm T}
        	-\mathbf A\mathbf C_x(\tau)
        	\label{eq:si-fdr-violation}
        \end{equation}
        quantifies the violation of the equilibrium
        fluctuation--dissipation relation.
        
        The short-time asymmetry defined in
        Eq.~\eqref{eq:short-time-asymmetry} satisfies
        \(\boldsymbol{\Omega}
        =(\mathbf C_0\mathbf A^{\mathrm T}-\mathbf A\mathbf C_0)/2\).
        Substituting the lagged propagation into
        Eq.~\eqref{eq:si-fdr-violation} therefore gives
        \begin{equation}
        	\boldsymbol{\Delta}(\tau)
        	=
        	e^{-\mathbf A\tau}
        	\bigl(\mathbf C_0\mathbf A^{\mathrm T}
        	-\mathbf A\mathbf C_0\bigr)
        	=
        	2e^{-\mathbf A\tau}\boldsymbol{\Omega}.
        	\label{eq:si-fdr-propagated-omega}
        \end{equation}
        Thus, the violation term at finite lag is determined by the
        propagation of \(\boldsymbol{\Omega}\) under the unperturbed
        network dynamics.
        Substituting its mode decomposition from
        Eq.~\eqref{eq:modal-time-asymmetry} yields
        \begin{equation}
        	\boldsymbol{\Delta}(\tau)
        	=
        	2\sum_{i,j=1}^{N}
        	\frac{\lambda_j^*-\lambda_i}
        	{\lambda_i+\lambda_j^*}
        	\Gamma_{ij}e^{-\lambda_i\tau}
        	\mathbf U_i\mathbf U_j^\dagger.
        	\label{eq:si-fdr-spectral}
        \end{equation}
        In the short-time limit,
        \begin{equation}
        	\lim_{\tau\to0^+}\boldsymbol{\Delta}(\tau)
        	=
        	2\boldsymbol{\Omega}.
        	\label{eq:si-fdr-zero-lag}
        \end{equation}
        At equilibrium, \(\boldsymbol{\Omega}=\mathbf0\), so the
        violation term vanishes at every positive lag.
        Equation~\eqref{eq:si-fdr-generalized} then reduces to
        the equilibrium fluctuation--dissipation relation,
        \(\partial_\tau\mathbf C_x(\tau)
        =-\mathbf R(\tau)\mathbf D\).
		
		\section{Heat currents at steady state}
		\label{app:heat-currents}
		
		We use units in which the common mobility of all nodes is unity.
		Following stochastic energetics~\cite{Sekimoto1998,Seifert2005,Harada2005},
		the mean heat current from node \(m\) into its environment
		at steady state is
		\begin{equation}
			J_m
			=
			-\langle(\mathbf A\mathbf x)_m\circ\dot x_m\rangle_0,
			\label{eq:si-heat-definition}
		\end{equation}
		where \(\langle\cdot\rangle_0\) denotes a steady-state average
		and \(\circ\) denotes the Stratonovich product.
		To evaluate this average, we first convert the stochastic
		integral to the It\^o convention.
		Over a short interval \(h>0\), the noise increment
		\(\Delta w_n=\int_t^{t+h}\eta_n(s)\,\mathrm ds\)
		satisfies
		\begin{equation}
			\begin{aligned}
				\langle\Delta w_n\Delta w_m\rangle
				&=
				\int_t^{t+h}\!\mathrm ds
				\int_t^{t+h}\!\mathrm ds'\,
				2D_{nm}\delta(s-s')\\
				&=2D_{nm}h.
			\end{aligned}
			\label{eq:si-heat-noise-increment}
		\end{equation}
		Drift increments are of order \(h\), whereas noise increments
		have typical magnitude of order \(h^{1/2}\).
		Only the noise therefore contributes to the quadratic
		covariation, giving
		\(\mathrm dx_n\,\mathrm dx_m=2D_{nm}\,\mathrm dt\).
		For the linear integrand \((\mathbf A\mathbf x)_m\),
		the Stratonovich-to-It\^o conversion becomes
		\begin{equation}
			\begin{aligned}
				(\mathbf A\mathbf x)_m\circ\mathrm dx_m
				&=
				(\mathbf A\mathbf x)_m\,\mathrm dx_m
				+\frac12\sum_n A_{mn}\,\mathrm dx_n\,\mathrm dx_m\\
				&=
				(\mathbf A\mathbf x)_m\,\mathrm dx_m
				+(\mathbf A\mathbf D)_{mm}\,\mathrm dt,
			\end{aligned}
			\label{eq:si-heat-conversion}
		\end{equation}
		where the integral on the right is interpreted in the It\^o
		sense.
		The noise increment has zero conditional mean given
		\(\mathbf x(t)\), so the Langevin equation implies
		\begin{equation}
			\langle(\mathbf A\mathbf x)_m\,\mathrm dx_m\rangle_0
			=
			-\langle(\mathbf A\mathbf x)_m^2\rangle_0\,\mathrm dt.
			\label{eq:si-heat-ito-average}
		\end{equation}
		Combining Eqs.~\eqref{eq:si-heat-conversion}
		and \eqref{eq:si-heat-ito-average} with the definition of
		\(J_m\) gives
		\begin{equation}
			\begin{aligned}
				J_m
				&=
				\langle(\mathbf A\mathbf x)_m^2\rangle_0
				-(\mathbf A\mathbf D)_{mm}\\
				&=
				\bigl(\mathbf A\mathbf C_0\mathbf A^{\mathrm T}
				-\mathbf A\mathbf D\bigr)_{mm},
			\end{aligned}
			\label{eq:si-heat-covariance}
		\end{equation}
		where \(\mathbf C_0=\langle\mathbf x\mathbf x^{\mathrm T}\rangle_0\)
		is the stationary covariance matrix.
		The Lyapunov equation and the definition of
		\(\boldsymbol{\Omega}\) imply
		\(\boldsymbol{\Omega}
		=\mathbf C_0\mathbf A^{\mathrm T}-\mathbf D\).
		The heat currents can therefore be collected into the vector
		\(\mathbf J=(J_1,\ldots,J_N)^{\mathrm T}\) as
		\begin{equation}
			\mathbf J
			=
			\operatorname{diag}(\mathbf A\boldsymbol{\Omega}),
			\label{eq:si-heat-omega}
		\end{equation}
		where \(\operatorname{diag}\) extracts the diagonal as a vector.
		Substituting the mode decomposition of
		\(\boldsymbol{\Omega}\) from
		Eq.~\eqref{eq:modal-time-asymmetry} and using
		\(\mathbf A\mathbf U_i=\lambda_i\mathbf U_i\)
		then recovers the heat current decomposition in
		Eq.~\eqref{eq:heat-spectral}.
		
		\section{Laplacian structure of the nonequilibrium response operators}
		\label{app:response-operators}
		
		For the stable reciprocal networks considered here,
		\(\mathbf U\) is real and orthogonal, and all eigenvalues
		satisfy \(\lambda_i>0\).
		Since \(\kappa_{ij}^{\alpha}=\kappa_{ji}^{\alpha}\)
		and \(\kappa_{ii}^{\alpha}=0\), the mode decomposition
		in Eq.~\eqref{eq:theta-spectral} can be written as
		\begin{equation}
			\Theta_{mn}^{\alpha}
			=\frac12\sum_{i,j=1}^{N}
			\kappa_{ij}^{\alpha}
			U_{mi}U_{ni}U_{mj}U_{nj},
			\label{eq:si-theta-entries}
		\end{equation}
		where \(\alpha\in\{\Omega,\mathrm J\}\).
		This expression is invariant under interchange of \(m\)
		and \(n\), establishing symmetry.
		Orthogonality also gives
		\begin{equation}
			\sum_{n=1}^{N}\Theta_{mn}^{\alpha}
			=\sum_{i<j}\kappa_{ij}^{\alpha}U_{mi}U_{mj}
			\sum_{n=1}^{N}U_{ni}U_{nj}=0,
			\label{eq:si-theta-row-sum}
		\end{equation}
		because the inner sum vanishes for distinct modes.
		Both operators therefore have zero row sums.
		It remains to establish that their off-diagonal entries
		are nonpositive.
		The spectral weights satisfy
		\begin{equation}
			\begin{aligned}
				\kappa_{ij}^{\mathrm J}
				&=\lambda_i+\lambda_j
				-\frac{4\lambda_i\lambda_j}{\lambda_i+\lambda_j},\\
				\kappa_{ij}^{\Omega}
				&=2-\frac{8\lambda_i\lambda_j}
				{(\lambda_i+\lambda_j)^2}.
			\end{aligned}
			\label{eq:si-weight-identities}
		\end{equation}
		For \(m\neq n\), orthogonality gives
		\(\sum_i U_{mi}U_{ni}=0\).
		Upon substitution into Eq.~\eqref{eq:si-theta-entries},
		the terms involving \(\lambda_i+\lambda_j\) and the
		constant term vanish, since their sums contain this
		zero factor.
		The remaining terms admit integral representations through
		the network propagator
		\(\mathbf G(t)=e^{-\mathbf A t}\), for which
		\[
		(\mathbf A\mathbf G(t))_{mn}
		=\sum_i\lambda_iU_{mi}U_{ni}e^{-\lambda_i t}.
		\]
		Using
		\(\int_0^\infty e^{-st}\,\mathrm dt=1/s\) and
		\(\int_0^\infty t e^{-st}\,\mathrm dt=1/s^2\)
		for \(s>0\), we obtain, for \(m\neq n\),
		\begin{equation}
			\begin{aligned}
				\Theta_{mn}^{\mathrm J}
				&=-2\sum_{i,j=1}^{N}
				\frac{\lambda_i\lambda_j}{\lambda_i+\lambda_j}
				U_{mi}U_{ni}U_{mj}U_{nj}\\
				&=-2\int_0^\infty
				\bigl[(\mathbf A\mathbf G(t))_{mn}\bigr]^2
				\,\mathrm dt\leq0,
			\end{aligned}
			\label{eq:si-theta-J-offdiagonal}
		\end{equation}
		and
		\begin{equation}
			\begin{aligned}
				\Theta_{mn}^{\Omega}
				&=-4\sum_{i,j=1}^{N}
				\frac{\lambda_i\lambda_j}{(\lambda_i+\lambda_j)^2}
				U_{mi}U_{ni}U_{mj}U_{nj}\\
				&=-4\int_0^\infty t\,
				\bigl[(\mathbf A\mathbf G(t))_{mn}\bigr]^2
				\,\mathrm dt\leq0.
			\end{aligned}
			\label{eq:si-theta-Omega-offdiagonal}
		\end{equation}
		Combining these inequalities with the zero row sums gives
		\begin{equation}
			\Theta_{mm}^{\alpha}
			=-\sum_{n\neq m}\Theta_{mn}^{\alpha}\geq0.
			\label{eq:si-theta-diagonal}
		\end{equation}
		Thus, each response operator is a symmetric weighted
		Laplacian with nonnegative edge weights
		\(-\Theta_{mn}^{\alpha}\) for \(m\neq n\).
		For any \(\mathbf z\in\mathbb R^N\), its quadratic form is
		\begin{equation}
			\mathbf z^{\mathrm T}\boldsymbol{\Theta}^{\alpha}\mathbf z
			=-\sum_{m<n}\Theta_{mn}^{\alpha}(z_m-z_n)^2
			\geq0,
			\label{eq:si-laplacian-quadratic}
		\end{equation}
		which establishes positive semidefiniteness.
		
		\section{Quadratic forms of the global nonequilibrium measures}
		\label{app:quadratic-forms}
		
		For reciprocal networks with independent local driving
		\(\mathbf D=\operatorname{diag}(\mathbf d)\), where \(d_m>0\),
		the diffusion matrix in mode coordinates is
		\(\boldsymbol{\Gamma}=\mathbf U^{\mathrm T}\mathbf D\mathbf U\),
		with
		\(\Gamma_{ij}
		=(\mathbf U_i\odot\mathbf U_j)^{\mathrm T}\mathbf d\).
		Using the orthogonal invariance of the Frobenius norm and
		Eq.~\eqref{eq:modal-time-asymmetry}, the global asymmetry
		intensity can be written as
		\begin{equation}
			\omega=\|\mathbf U^{\mathrm T}\boldsymbol{\Omega}\mathbf U\|_{\mathrm F}^{2}=2\sum_{i<j}\frac{(\lambda_i-\lambda_j)^2}{(\lambda_i+\lambda_j)^2}\Gamma_{ij}^{2}=\mathbf d^{\mathrm T}\boldsymbol{\Theta}^{\Omega}\mathbf d.
			\label{eq:si-omega-quadratic}
		\end{equation}
		This quadratic form is exact for arbitrary positive
		driving intensities.
		The entropy production rate follows from the heat currents
		into the reservoirs.
		Pairing the \((i,j)\) and \((j,i)\) contributions in
		Eq.~\eqref{eq:heat-spectral} gives
		\begin{equation}
			\mathbf J=-\sum_{i<j}\frac{(\lambda_i-\lambda_j)^2}{\lambda_i+\lambda_j}\Gamma_{ij}(\mathbf U_i\odot\mathbf U_j)=-\boldsymbol{\Theta}^{\mathrm J}\mathbf d.
			\label{eq:si-heat-driving}
		\end{equation}
		Let \(T_m\) denote the temperature of reservoir \(m\).
		Since \(J_m\) is the mean heat current into this reservoir,
		its entropy flow rate is
		\begin{equation}
			\dot S_m^{\mathrm{env}}=\frac{J_m}{T_m}.
			\label{eq:si-reservoir-entropy-flow}
		\end{equation}
		For a common mobility \(\mu\), the Einstein relation connects
		the driving intensity to the reservoir temperature,
		\begin{equation}
			d_m=\mu k_{\mathrm B}T_m.
			\label{eq:si-einstein-relation}
		\end{equation}
		In the units adopted here, \(\mu=k_{\mathrm B}=1\), so
		\(d_m=T_m\).
		At steady state, \(\dot S_{\mathrm{sys}}=0\), and the total
		entropy production rate equals the entropy flow rate into
		the reservoirs~\cite{Seifert2012},
		\begin{equation}
			\sigma=\dot S_{\mathrm{sys}}+\sum_m\dot S_m^{\mathrm{env}}=\sum_m\frac{J_m}{T_m}=\sum_m\frac{J_m}{d_m}.
			\label{eq:si-steady-entropy-balance}
		\end{equation}
		To express \(\sigma\) in terms of differences between local
		driving intensities, we use the zero row sums of
		\(\boldsymbol{\Theta}^{\mathrm J}\), established in
		Appendix~\ref{app:response-operators},
		\begin{equation}
			\begin{aligned}
				(\boldsymbol{\Theta}^{\mathrm J}\mathbf d)_m
				&=\Theta_{mm}^{\mathrm J}d_m
				+\sum_{n\neq m}\Theta_{mn}^{\mathrm J}d_n\\
				&=\sum_{n\neq m}
				\Theta_{mn}^{\mathrm J}(d_n-d_m).
			\end{aligned}
			\label{eq:si-heat-driving-differences}
		\end{equation}
		Substituting \(\mathbf J=-\boldsymbol{\Theta}^{\mathrm J}\mathbf d\)
		into Eq.~\eqref{eq:si-steady-entropy-balance} and combining
		paired terms using
		\(\Theta_{mn}^{\mathrm J}=\Theta_{nm}^{\mathrm J}\) yield
		\begin{equation}
			\begin{aligned}
				\sigma
				&=-\sum_m
				\frac{(\boldsymbol{\Theta}^{\mathrm J}\mathbf d)_m}{d_m}\\
				&=-\sum_m\sum_{n\neq m}
				\Theta_{mn}^{\mathrm J}\frac{d_n-d_m}{d_m}\\
				&=-\sum_{m<n}\Theta_{mn}^{\mathrm J}
				\left(\frac{d_n}{d_m}+\frac{d_m}{d_n}-2\right)\\
				&=-\sum_{m<n}\Theta_{mn}^{\mathrm J}
				\frac{(d_m-d_n)^2}{d_md_n}.
			\end{aligned}
			\label{eq:si-epr-heat}
		\end{equation}
		This expression is exact for positive driving intensities.
		Applying the Laplacian identity,
		Eq.~\eqref{eq:si-laplacian-quadratic}, to
		Eq.~\eqref{eq:si-omega-quadratic} gives the corresponding
		pairwise expression for \(\omega\).
		Together, these expressions recover
		Eq.~\eqref{eq:global-irreversibility}.
		The quadratic approximation for \(\sigma\) is obtained by
		expanding about uniform unit driving,
		\(\mathbf d=\mathbf1+\delta\mathbf d\).
		For \(i\neq j\), orthogonality implies
		\((\mathbf U_i\odot\mathbf U_j)^{\mathrm T}\mathbf1
		=\mathbf U_i^{\mathrm T}\mathbf U_j=0\), and hence
		\begin{equation}
			\Gamma_{ij}
			=(\mathbf U_i\odot\mathbf U_j)^{\mathrm T}\mathbf d
			=(\mathbf U_i\odot\mathbf U_j)^{\mathrm T}\delta\mathbf d.
			\label{eq:si-driving-projection}
		\end{equation}
		The zero row sums of the response operators also give
		\(\omega=\delta\mathbf d^{\mathrm T}
		\boldsymbol{\Theta}^{\Omega}\delta\mathbf d\) and
		\(\mathbf J=-\boldsymbol{\Theta}^{\mathrm J}\delta\mathbf d\).
		Near equilibrium, \(\|\delta\mathbf d\|_2\ll1\), we expand
		\((1+\delta d_m)^{-1}
		=1-\delta d_m+O(\delta d_m^2)\).
		The linear contribution to \(\sigma\) vanishes because
		symmetry and zero row sums imply
		\(\mathbf1^{\mathrm T}\boldsymbol{\Theta}^{\mathrm J}
		=\mathbf0^{\mathrm T}\).
		The leading contribution is therefore quadratic,
		\begin{equation}
			\begin{aligned}
				\sigma
				&=-\sum_m
				\frac{(\boldsymbol{\Theta}^{\mathrm J}
					\delta\mathbf d)_m}{1+\delta d_m}\\
				&=\delta\mathbf d^{\mathrm T}
				\boldsymbol{\Theta}^{\mathrm J}\delta\mathbf d
				+O(\|\delta\mathbf d\|_2^3)\\
				&=\sum_{i<j}\kappa_{ij}^{\mathrm J}\Gamma_{ij}^{2}
				+O(\|\delta\mathbf d\|_2^3).
			\end{aligned}
			\label{eq:si-sigma-quadratic}
		\end{equation}
		The last equality follows from
		Eqs.~\eqref{eq:theta-spectral}
		and \eqref{eq:si-driving-projection}.
		Thus, \(\boldsymbol{\Theta}^{\mathrm J}\) determines the
		leading-order entropy production rate near equilibrium.
		
		\section{Bounds on global asymmetry intensity per unit
			entropy production rate}
		\label{app:generalized-eigenvalue-bounds}
		
		For weakly heterogeneous driving
		\(\mathbf d=\mathbf1+\delta\mathbf d\), with
		\(\delta\mathbf d\perp\mathbf1\) and
		\(\|\delta\mathbf d\|_2\ll1\),
		Appendix~\ref{app:quadratic-forms} gives
		\begin{equation}
			\begin{aligned}
				\omega
				&=\delta\mathbf d^{\mathrm T}
				\boldsymbol{\Theta}^{\Omega}\delta\mathbf d,\\
				\sigma
				&=\sigma_2+O(\|\delta\mathbf d\|_2^3),
				\qquad
				\sigma_2=\delta\mathbf d^{\mathrm T}
				\boldsymbol{\Theta}^{\mathrm J}\delta\mathbf d.
			\end{aligned}
			\label{eq:si-ratio-expansion}
		\end{equation}
		The expression for \(\omega\) is exact, while \(\sigma_2\)
		is the leading quadratic contribution to the entropy
		production rate near equilibrium.
		We first establish that the ratio \(\omega/\sigma_2\)
		is well defined for every nonzero driving perturbation
		in \(\mathbf1^\perp\).
		For \(m\neq n\) with \(A_{mn}\neq0\), continuity of
		\((\mathbf A\mathbf G(t))_{mn}\) and its nonzero value
		at \(t=0\) ensure that the integral representations in
		Eqs.~\eqref{eq:si-theta-J-offdiagonal}
		and \eqref{eq:si-theta-Omega-offdiagonal} give
		\(\Theta_{mn}^{\alpha}<0\).
		Thus, every direct interaction generates an edge in both
		response networks.
		For a connected interaction network, both response networks
		are therefore connected.
		Their Laplacian structure then ensures that the uniform
		vector is the only zero mode and that both operators
		are positive definite on \(\mathbf1^\perp\).
		On this subspace, consider the generalized eigenvalue problem
		\begin{equation}
			\boldsymbol{\Theta}^{\Omega}\boldsymbol{\ell}_r
			=\rho_r\boldsymbol{\Theta}^{\mathrm J}\boldsymbol{\ell}_r,
			\qquad
			\boldsymbol{\ell}_r^{\mathrm T}\mathbf1=0,
			\label{eq:si-generalized-eigenvectors}
		\end{equation}
		with \(r=1,\ldots,N-1\).
		Symmetry and positive definiteness guarantee a complete
		generalized eigenbasis, which can be normalized as
		\(\boldsymbol{\ell}_r^{\mathrm T}
		\boldsymbol{\Theta}^{\mathrm J}\boldsymbol{\ell}_s
		=\delta_{rs}\), where \(\delta_{rs}\) is the Kronecker delta.
		Any driving perturbation in \(\mathbf1^\perp\) can then
		be expanded as
		\begin{equation}
			\delta\mathbf d
			=\sum_{r=1}^{N-1}c_r\boldsymbol{\ell}_r,
			\qquad
			c_r=\boldsymbol{\ell}_r^{\mathrm T}
			\boldsymbol{\Theta}^{\mathrm J}\delta\mathbf d.
			\label{eq:si-driving-expansion}
		\end{equation}
		The generalized eigenvalue equation also gives
		\(\boldsymbol{\ell}_r^{\mathrm T}
		\boldsymbol{\Theta}^{\Omega}\boldsymbol{\ell}_s
		=\rho_s\delta_{rs}\).
		Substitution into the two quadratic forms therefore yields
		\begin{equation}
			\omega=\sum_r\rho_r c_r^2,
			\qquad
			\sigma_2=\sum_r c_r^2.
			\label{eq:si-eigenvector-quadratic}
		\end{equation}
		Let \(\rho_{\min}\) and \(\rho_{\max}\) denote the
		smallest and largest generalized eigenvalues.
		Since \(c_r^2\geq0\),
		\begin{equation}
			\rho_{\min}\sum_r c_r^2
			\leq\sum_r\rho_r c_r^2
			\leq\rho_{\max}\sum_r c_r^2.
			\label{eq:si-weighted-eigenvalue-bounds}
		\end{equation}
		For nonzero \(\delta\mathbf d\), positive definiteness
		ensures \(\sigma_2=\sum_r c_r^2>0\).
		Dividing by \(\sigma_2\) gives
		\begin{equation}
			\rho_{\min}
			\leq\frac{\omega}{\sigma_2}
			=\frac{\sum_r\rho_r c_r^2}{\sum_r c_r^2}
			\leq\rho_{\max}.
			\label{eq:si-generalized-rayleigh}
		\end{equation}
		Thus, the ratio is a weighted average of the generalized
		eigenvalues and lies between their minimum and maximum.
		These bounds are attainable.
		For a fixed and sufficiently small quadratic entropy
		production rate \(\sigma_2=\sigma_0>0\), the driving
		perturbation
		\(\delta\mathbf d=\sqrt{\sigma_0}\boldsymbol{\ell}_{\max}\)
		satisfies the constraint and gives
		\begin{equation}
			\omega_{\max}
			=\sigma_0\boldsymbol{\ell}_{\max}^{\mathrm T}
			\boldsymbol{\Theta}^{\Omega}\boldsymbol{\ell}_{\max}
			=\rho_{\max}\sigma_0.
			\label{eq:si-optimal-driving}
		\end{equation}
		Similarly,
		\(\delta\mathbf d=\sqrt{\sigma_0}\boldsymbol{\ell}_{\min}\)
		attains the lower bound
		\(\omega_{\min}=\rho_{\min}\sigma_0\).
		Finally, positive definiteness of
		\(\boldsymbol{\Theta}^{\mathrm J}\) on \(\mathbf1^\perp\)
		bounds \(\sigma_2\) below by a positive multiple of
		\(\|\delta\mathbf d\|_2^2\).
		The cubic correction in Eq.~\eqref{eq:si-ratio-expansion}
		therefore produces a relative error of order
		\(\|\delta\mathbf d\|_2\), giving
		\begin{equation}
			\frac{\omega}{\sigma}
			=\frac{\omega}{\sigma_2}
			+O(\|\delta\mathbf d\|_2).
			\label{eq:si-ratio-correction}
		\end{equation}
		The generalized eigenvalues consequently bound
		\(\omega/\sigma\) to leading order near equilibrium.
		The ratio approaches \(\rho_{\min}\) or \(\rho_{\max}\)
		as the driving amplitude tends to zero along
		\(\boldsymbol{\ell}_{\min}\) or
		\(\boldsymbol{\ell}_{\max}\), respectively.
		
		\section{Fiedler modes and spectral bounds in ideal hierarchical modular networks}
		\label{app:fiedler-extrema}
		
		Consider an ideal \(\mathrm H_3\) network with \(N=2n\) nodes
		divided into two equally sized largest modules.
		Every pair of nodes in different modules is connected with
		weight \(b>0\), while every pair of distinct nodes within either module is connected with weight greater than \(b\).
		The normalized uniform and Fiedler modes are
		\begin{equation}
			\mathbf U_1=\frac{1}{\sqrt N}
			\begin{pmatrix}
				\mathbf1_n\\
				\mathbf1_n
			\end{pmatrix},
			\qquad
			\mathbf U_2=\frac{1}{\sqrt N}
			\begin{pmatrix}
				\mathbf1_n\\
				-\mathbf1_n
			\end{pmatrix}.
			\label{eq:app-fiedler-modes}
		\end{equation}
		Since \(\mathbf U_2\) is constant within each module,
		only intermodule connections contribute to
		\(\mathbf L\mathbf U_2=Nb\,\mathbf U_2\).
		Relative to a complete graph with edge weight \(b\),
		the additional intramodule weights raise all remaining
		nonuniform eigenvalues.
		Thus, \(k_2=Nb\) and \(k_j>k_2\) for \(j>2\).
		For \(\mathbf A=\gamma\mathbf I+g\mathbf L\), the corresponding
		relaxation rates are
		\(\lambda_1=\gamma\) and \(\lambda_2=\gamma+gk_2\).
		We first derive bounds valid for arbitrary nonuniform driving
		and then show that driving along \(\mathbf U_2\) attains them.
		The relevant spectral edges have the variational expressions
		\begin{equation}
			\begin{aligned}
				\mu_{\min+}^{\Omega}
				&=\min_{\delta\mathbf d}
				\frac{\omega}{\|\delta\mathbf d\|_2^2},\\
				\mu_{\min+}^{\mathrm J}
				&=\min_{\delta\mathbf d}
				\frac{\sigma_2}{\|\delta\mathbf d\|_2^2},\\
				\rho_{\max}
				&=\max_{\delta\mathbf d}
				\frac{\omega}{\sigma_2},
			\end{aligned}
			\label{eq:app-fiedler-variational}
		\end{equation}
		where the extrema are taken over nonzero
		\(\delta\mathbf d\perp\mathbf1\).
		For \(\mathbf d=\mathbf1+\delta\mathbf d\),
		the uniform mode \(\mathbf U_1=\mathbf1/\sqrt N\) gives
		\(\Gamma_{1j}
		=\mathbf U_j^{\mathrm T}\delta\mathbf d/\sqrt N\)
		for \(j>1\).
		Orthonormality and \(\delta\mathbf d\perp\mathbf1\) therefore imply
		\begin{equation}
			\Gamma_{1j}
			=\frac{\mathbf U_j^{\mathrm T}\delta\mathbf d}{\sqrt N},
			\qquad
			\sum_{j=2}^{N}\Gamma_{1j}^{2}
			=\frac{\|\delta\mathbf d\|_2^2}{N}.
			\label{eq:app-uniform-mode-correlations}
		\end{equation}
		For both \(\alpha=\Omega\) and \(\alpha=\mathrm J\),
		the weights \(\kappa_{1j}^{\alpha}\) increase with
		\(\lambda_j\geq\lambda_1\), so
		\(\kappa_{1j}^{\alpha}\geq\kappa_{12}^{\alpha}\)
		for \(j>1\).
		Retaining only the nonnegative contributions involving
		the uniform mode consequently gives
		\begin{equation}
			\begin{aligned}
				\omega
				&\geq\sum_{j=2}^{N}
				\kappa_{1j}^{\Omega}\Gamma_{1j}^{2}
				\geq\frac{\kappa_{12}^{\Omega}}{N}
				\|\delta\mathbf d\|_2^2,\\
				\sigma_2
				&\geq\sum_{j=2}^{N}
				\kappa_{1j}^{\mathrm J}\Gamma_{1j}^{2}
				\geq\frac{\kappa_{12}^{\mathrm J}}{N}
				\|\delta\mathbf d\|_2^2.
			\end{aligned}
			\label{eq:app-fiedler-lower-bounds}
		\end{equation}
		An upper bound on the ratio follows from
		\(\kappa_{ij}^{\Omega}
		=2\kappa_{ij}^{\mathrm J}/(\lambda_i+\lambda_j)\),
		\begin{equation}
			\frac{\omega}{\sigma_2}
			=
			\frac{\sum_{i<j}\frac{2}{\lambda_i+\lambda_j}
				\kappa_{ij}^{\mathrm J}\Gamma_{ij}^{2}}
			{\sum_{i<j}\kappa_{ij}^{\mathrm J}\Gamma_{ij}^{2}}
			\leq\frac{2}{\lambda_1+\lambda_2}.
			\label{eq:app-fiedler-ratio-bound}
		\end{equation}
		The ratio is a weighted average of
		\(2/(\lambda_i+\lambda_j)\) with nonnegative weights,
		and \(\lambda_1+\lambda_2\) is the smallest sum for
		distinct modes.
		Since these inequalities hold for every nonzero
		\(\delta\mathbf d\perp\mathbf1\), the variational expressions
		in Eq.~\eqref{eq:app-fiedler-variational} yield
		\begin{equation}
			\mu_{\min+}^{\Omega}\geq\frac{\kappa_{12}^{\Omega}}{N},
			\quad
			\mu_{\min+}^{\mathrm J}\geq\frac{\kappa_{12}^{\mathrm J}}{N},
			\quad
			\rho_{\max}\leq\frac{2}{\lambda_1+\lambda_2}.
			\label{eq:app-fiedler-eigenvalue-bounds}
		\end{equation}
		To establish attainability, choose
		\(\delta\mathbf d=a\mathbf U_2\) with nonzero and
		sufficiently small \(a\).
		The driving intensities are then \(1+a/\sqrt N\) in one
		module and \(1-a/\sqrt N\) in the other.
		Uniform intermodule coupling allows the remaining relaxation
		modes to be chosen with support in either module and zero
		sum within it.
		Because \(\mathbf D\) is constant within each module,
		these modes are also eigenvectors of \(\mathbf D\).
		All off-diagonal entries of \(\boldsymbol{\Gamma}\)
		involving these modes therefore vanish, leaving only
		\(\Gamma_{12}=\Gamma_{21}=a/\sqrt N\).
		Thus, only the pair \((1,2)\) contributes to
		\(\omega\) and \(\sigma_2\).
		Using \(\|\delta\mathbf d\|_2^2=a^2\), we obtain
		\begin{equation}
			\frac{\omega}{\|\delta\mathbf d\|_2^2}
			=\frac{\kappa_{12}^{\Omega}}{N},
			\quad
			\frac{\sigma_2}{\|\delta\mathbf d\|_2^2}
			=\frac{\kappa_{12}^{\mathrm J}}{N},
			\quad
			\frac{\omega}{\sigma_2}
			=\frac{2}{\lambda_1+\lambda_2}.
			\label{eq:app-fiedler-saturation}
		\end{equation}
		These values saturate all three bounds in
		Eq.~\eqref{eq:app-fiedler-eigenvalue-bounds}.
		Substituting \(\lambda_1=\gamma\) and
		\(\lambda_2=\gamma+gk_2\) therefore gives the exact spectral edges,
		\begin{equation}
			\begin{aligned}
				\mu_{\min+}^{\Omega}
				&=\frac{\kappa_{12}^{\Omega}}{N}
				=\frac{2(gk_2)^2}{N(2\gamma+gk_2)^2},\\
				\mu_{\min+}^{\mathrm J}
				&=\frac{\kappa_{12}^{\mathrm J}}{N}
				=\frac{(gk_2)^2}{N(2\gamma+gk_2)},\\
				\rho_{\max}
				&=\frac{2}{2\gamma+gk_2}.
			\end{aligned}
			\label{eq:app-fiedler-spectral-edges}
		\end{equation}
		Driving along \(\mathbf U_2\) therefore simultaneously
		minimizes \(\omega\) and \(\sigma_2\) at fixed driving
		heterogeneity and maximizes \(\omega/\sigma_2\).
		The three extrema share the driving direction
		\(\delta\mathbf d
		\parallel\mathbf v_{\min+}^{\Omega}
		\parallel\mathbf v_{\min+}^{\mathrm J}
		\parallel\boldsymbol{\ell}_{\max}
		\parallel\mathbf U_2\).
		The same argument extends to ideal \(\mathrm H_2\) and
		\(\mathrm H_1\) networks with equally sized largest retained
		modules, uniform intermodule coupling, and stronger
		intramodule connections.
		Driving along a Fiedler mode likewise attains all three
		bounds in Eq.~\eqref{eq:app-fiedler-spectral-edges}.
		
		\section{Empirical covariance estimation and inference of the diffusion matrix}
		\label{app:driving-inference}
		
		\subsection{Empirical covariance estimation and time-reversal asymmetry}
		
		For each participant, the four resting-state runs are
		concatenated to form a time series
		\(\mathbf x(t)\in\mathbb R^N\) with \(N=360\) brain regions
		and \(T=4800\) time points, where \(t\) is measured in units
		of the repetition time (\(\mathrm{TR}=0.72\,\mathrm{s}\)).
		The regional signals are demeaned and rescaled by a common
		factor to set the mean regional sample variance to unity
		while preserving regional variance ratios.
		To ensure consistent sampling across lags, all covariance
		estimates use the same \(M=T-K\) starting time points,
		where \(K\) is the maximum lag considered.
		For \(k=0,\ldots,K\), the empirical covariance is
		\begin{equation}
			\mathbf C_k^{\mathrm{emp}}
			=\frac{1}{M-1}\sum_{t=1}^{M}
			\mathbf x(t+k)\mathbf x(t)^{\mathrm T}.
			\label{eq:H-empirical-covariance}
		\end{equation}
		These matrices represent the unnormalized cross-correlations
		of the demeaned signals.
		At positive lags, their asymmetry is
		\begin{equation}
			\mathbf B_k=\mathbf C_k-\mathbf C_k^{\mathrm T}.
			\label{eq:H-covariance-asymmetry}
		\end{equation}
		The same notation applies to the empirical, model, and basis
		matrices used below.
		The short-time expansion
		\(\mathbf B(\tau)=2\tau\boldsymbol{\Omega}+O(\tau^2)\)
		shows that \(\mathbf B(\tau)/(2\tau)\) approaches
		\(\boldsymbol{\Omega}\) as \(\tau\to0^+\).
		This motivates a finite-lag counterpart of the global
		asymmetry intensity
		\(\omega=\|\boldsymbol{\Omega}\|_{\mathrm F}^{2}\).
		For each participant, it is estimated at \(\tau=k>0\) as
		\begin{equation}
			\widehat{\omega}(\tau)
			=\left\|\frac{\mathbf B_k^{\mathrm{emp}}}{2\tau}
			\right\|_{\mathrm F}^{2}
			=\frac{\left\|\mathbf B_k^{\mathrm{emp}}\right\|_{\mathrm F}^{2}}
			{4\tau^{2}}.
			\label{eq:H-finite-lag-asymmetry}
		\end{equation}
		To assess the nonequilibrium character of resting-state brain
		activity, the empirical global asymmetry intensity is compared
		with that of temporally shuffled time series.
		For each participant and \(p\in\{0.25,0.5,1\}\),
		\(pT\) time points are selected uniformly without replacement.
		The corresponding whole-brain activity vectors are randomly
		shuffled among the selected positions, leaving all other
		positions unchanged.
		All brain regions therefore share the same temporal shuffle,
		preserving the set of instantaneous spatial activity patterns
		while disrupting their temporal order.
		The case \(p=1\) corresponds to shuffling the entire time series.
		One shuffled series is generated per participant at each \(p\),
		and its lagged covariances and global asymmetry intensity
		are recomputed using the same estimators.
		Empirical and shuffled intensities are then averaged across
		participants at each lag.
		
		\subsection{Inference of the diffusion matrix}
		
		For each participant, we infer the regional driving intensities by
		fitting the model to the empirical zero-lag covariance and the
		asymmetry of lagged cross-correlations. The interaction matrix is constructed from
		the individual SC matrix \(\mathbf W\) through its graph Laplacian
		\(\mathbf L\) as
		\(\mathbf A(\gamma,g)=\gamma\mathbf I+g\mathbf L\).
		The driving intensities are optimized for each candidate pair
		\((\gamma,g)\), after which the pair with the smallest normalized
		fitting error is selected.
		For independent stochastic driving, the diffusion matrix is
		\(\mathbf D=\operatorname{diag}(\mathbf d)\), where \(d_i>0\)
		is the driving intensity in region \(i\). The model zero-lag and
		lagged covariances satisfy
		\begin{equation}
			\mathbf A\mathbf C_0^{\mathrm{mod}}
			+\mathbf C_0^{\mathrm{mod}}\mathbf A=2\mathbf D,\qquad
			\mathbf C_k^{\mathrm{mod}}=e^{-k\mathbf A}\mathbf C_0^{\mathrm{mod}}.
			\label{eq:H-model-covariance}
		\end{equation}
		To express their dependence on the driving intensities, we write
		\(\mathbf D=\sum_{i=1}^{N}d_i\mathbf E_i\), where the only nonzero
		entry of \(\mathbf E_i\) is \((E_i)_{ii}=1\).
		Unit driving in region \(i\) alone gives the covariance basis
		matrices \(\mathbf C_k^{(i)}\), which satisfy
		\begin{equation}
			\mathbf A\mathbf C_0^{(i)}+\mathbf C_0^{(i)}\mathbf A
			=2\mathbf E_i,\qquad
			\mathbf C_k^{(i)}=e^{-k\mathbf A}\mathbf C_0^{(i)}.
			\label{eq:H-basis-covariance}
		\end{equation}
		By linearity of the Lyapunov equation and lagged propagation,
		\begin{equation}
			\mathbf C_k^{\mathrm{mod}}=\sum_{i=1}^{N}d_i\mathbf C_k^{(i)},\qquad
			\mathbf B_k^{\mathrm{mod}}=\sum_{i=1}^{N}d_i\mathbf B_k^{(i)}.
			\label{eq:H-linear-covariance}
		\end{equation}
		For the model, empirical and basis matrices, we use the common
		notation \(\mathbf T_0=\mathbf C_0\) and
		\(\mathbf T_k=\mathbf B_k\) for \(k\geq1\).
		The fitting objective is the sum of squared Frobenius residuals,
		each normalized by the squared norm of its empirical target,
		\begin{equation}
			\mathcal L_{\mathcal T}(\gamma,g,\mathbf d)
			=\sum_{k\in\mathcal T}w_k
			\left\|\mathbf T_k^{\mathrm{mod}}(\gamma,g,\mathbf d)
			-\mathbf T_k^{\mathrm{emp}}\right\|_{\mathrm F}^{2},
			\label{eq:H-loss}
		\end{equation}
		where \(\mathcal T\) is the set of fitted lags
		(e.g., \(\mathcal T_{01}=\{0,1\}\) for a joint fit to
		\(\mathbf C_0\) and \(\mathbf B_1\)) and
		\(w_k=\|\mathbf T_k^{\mathrm{emp}}\|_{\mathrm F}^{-2}\).
		Substituting Eq.~\eqref{eq:H-linear-covariance} into
		Eq.~\eqref{eq:H-loss} and using the Frobenius inner product
		\(\langle\mathbf X,\mathbf Y\rangle_{\mathrm F}
		=\operatorname{Tr}(\mathbf X^{\mathrm T}\mathbf Y)\) gives
		\begin{equation}
			\begin{aligned}
				\mathcal L_{\mathcal T}
				&=\sum_{k\in\mathcal T}w_k
				\left\|\sum_{i=1}^{N}d_i\mathbf T_k^{(i)}
				-\mathbf T_k^{\mathrm{emp}}\right\|_{\mathrm F}^{2}\\
				&=\mathbf d^{\mathrm T}\mathcal G_{\mathcal T}\mathbf d
				-2\mathbf z_{\mathcal T}^{\mathrm T}\mathbf d+\beta_{\mathcal T}.
			\end{aligned}
			\label{eq:H-quadratic-loss}
		\end{equation}
		with coefficients
		\begin{equation}
			\begin{aligned}
				[\mathcal G_{\mathcal T}]_{ij}
				&=\sum_{k\in\mathcal T}w_k
				\left\langle\mathbf T_k^{(i)},\mathbf T_k^{(j)}
				\right\rangle_{\mathrm F},\\
				[\mathbf z_{\mathcal T}]_i
				&=\sum_{k\in\mathcal T}w_k
				\left\langle\mathbf T_k^{(i)},\mathbf T_k^{\mathrm{emp}}
				\right\rangle_{\mathrm F},\\
				\beta_{\mathcal T}
				&=\sum_{k\in\mathcal T}w_k
				\left\|\mathbf T_k^{\mathrm{emp}}\right\|_{\mathrm F}^{2}.
			\end{aligned}
			\label{eq:H-quadratic-coefficients}
		\end{equation}
		Since
		\begin{equation}
			\mathbf y^{\mathrm T}\mathcal G_{\mathcal T}\mathbf y
			=\sum_{k\in\mathcal T}w_k
			\left\|\sum_{i=1}^{N}y_i\mathbf T_k^{(i)}
			\right\|_{\mathrm F}^{2}\geq0
			\label{eq:H-positive-semidefinite}
		\end{equation}
		for every \(\mathbf y\in\mathbb R^N\), the matrix
		\(\mathcal G_{\mathcal T}\) is positive semidefinite.
		For fixed \((\gamma,g)\), the optimal driving intensities therefore
		solve the convex quadratic program
		\begin{equation}
			\mathbf d_{\mathcal T}^{*}(\gamma,g)
			=\underset{\mathbf d>\mathbf0}{\operatorname{arg\,min}}
			\left[
			\tfrac12\mathbf d^{\mathrm T}\mathcal G_{\mathcal T}\mathbf d
			-\mathbf z_{\mathcal T}^{\mathrm T}\mathbf d
			\right].
			\label{eq:H-quadratic-program}
		\end{equation}
		The normalized fitting error at the fitted driving intensities is
		\begin{equation}
			\mathcal E_{\mathcal T}(\gamma,g)
			=\frac{\mathcal L_{\mathcal T}
				\bigl(\gamma,g,\mathbf d_{\mathcal T}^{*}(\gamma,g)\bigr)}
			{\beta_{\mathcal T}}.
			\label{eq:H-normalized-error}
		\end{equation}
		Minimizing this error over the search grid \(\mathcal P\) gives
		the optimal parameter pair and corresponding diffusion matrix,
		\begin{equation}
			\begin{aligned}
				(\gamma_{\mathcal T}^{*},g_{\mathcal T}^{*})
				&=\underset{(\gamma,g)\in\mathcal P}{\operatorname{arg\,min}}\,
				\mathcal E_{\mathcal T}(\gamma,g),\\
				\mathbf D_{\mathcal T}^{*}
				&=\operatorname{diag}\!\left[
				\mathbf d_{\mathcal T}^{*}
				(\gamma_{\mathcal T}^{*},g_{\mathcal T}^{*})\right].
			\end{aligned}
			\label{eq:H-optimal-parameters}
		\end{equation}
		For each participant, we solve Eq.~\eqref{eq:H-quadratic-program}
		at each grid point using the \texttt{interior-point-convex}
		algorithm in MATLAB's \texttt{quadprog}.
		The numerical optimization settings and fitting results
		are reported in the Supplemental Material.
	\bibliography{references}
\end{document}

\title{Supplementary Material for\\ Network topology imposes spectral constraints on collective nonequilibrium responses}
	
	\author{Dingjie Wu}
	\affiliation{
		State Key Laboratory for Strength and Vibration of Mechanical Structures,
		Xi'an Jiaotong University, Xi'an 710049, China
	}
	\affiliation{
		School of Aerospace Engineering, Xi'an Jiaotong University, Xi'an 710049, China
	}
	
	\author{Yaozu Wang}
	\affiliation{
		College of Intelligence Science and Technology, National University of Defense Technology, Changsha 410073, China
	}
	
	\author{Rong Wang}
	\email[Contact author: ]{tobous@mail.xjtu.edu.cn}
	\affiliation{
		State Key Laboratory for Strength and Vibration of Mechanical Structures,
		Xi'an Jiaotong University, Xi'an 710049, China
	}
	\affiliation{
		School of Aerospace Engineering, Xi'an Jiaotong University, Xi'an 710049, China
	}
	
	\author{Ying Wu}
	\email[Contact author: ]{wying36@xjtu.edu.cn}
	\affiliation{
		State Key Laboratory for Strength and Vibration of Mechanical Structures,
		Xi'an Jiaotong University, Xi'an 710049, China
	}
	\affiliation{
		School of Aerospace Engineering, Xi'an Jiaotong University, Xi'an 710049, China
	}
	
	\date{\today}
	\maketitle
	
	\appendix
	
	\setcounter{equation}{0}
	\renewcommand{\theequation}{A\arabic{equation}}
	
	\section{Network topology and equilibrium conditions}
	\label{sec:si-mode-correlations}
	
	We consider a real diagonalizable interaction matrix
	\(\mathbf A\) with \(\operatorname{Re}\lambda_i>0\)
	and a real symmetric positive definite diffusion matrix
	\(\mathbf D\).
	As in the main text, all variables are even under time reversal.
	Let \(R_i(\tau)\) and \(R_{ij}(\tau)\), with \(i\neq j\),
	denote the first and second ratios in Eq.~(7) of the main text.
	Equilibrium requires every defined ratio to equal unity
	for all \(\tau>0\).
	Since
	\(\boldsymbol{\Gamma}
	=\mathbf U^{-1}\mathbf D\mathbf U^{-\dagger}\succ0\),
	each diagonal entry \(\Gamma_{ii}\) is positive.
	Thus, \(R_i(\tau)\) is always defined and equals unity
	for all lags if and only if
	\(\operatorname{Im}\lambda_i=0\).
	For distinct modes, \(R_{ij}(\tau)\) is defined when
	\(\Gamma_{ij}\neq0\) and equals unity for all lags
	if and only if \(\lambda_i=\lambda_j^*\).
	These conditions distinguish the following four cases.
	\paragraph{Reciprocal networks.}
	For \(\mathbf A=\mathbf A^{\mathrm T}\), all eigenvalues
	are real and \(\mathbf U\) can be chosen real and orthogonal.
	Consequently, \(R_i(\tau)=1\) for every mode.
	For homogeneous independent driving,
	\(\mathbf D=\mathbf I\), the mode diffusion matrix is
	\(\boldsymbol{\Gamma}=\mathbf I\), so
	\(\Gamma_{ij}=0\) for \(i\neq j\) and the equilibrium
	conditions are satisfied.
	Spatially heterogeneous or correlated driving can generate
	off-diagonal entries of
	\(\boldsymbol{\Gamma}=\mathbf U^{\mathrm T}\mathbf D\mathbf U\).
	Correlated modes with different relaxation rates then satisfy
	\(R_{ij}(\tau)\neq1\) for \(\tau>0\).
	For example,
	\begin{equation}
		\mathbf A=
		\begin{pmatrix}
			1&0\\
			0&2
		\end{pmatrix},
		\qquad
		\mathbf D=
		\begin{pmatrix}
			1&0.5\\
			0.5&1
		\end{pmatrix}.
		\label{eq:si-mode-correlated-noise}
	\end{equation}
	Here, \(\lambda_1=1\), \(\lambda_2=2\), and
	\(\Gamma_{12}=\Gamma_{21}=0.5\), giving
	\(R_{12}(\tau)=e^\tau\neq1\) for \(\tau>0\).
	Spatially structured driving remains compatible with
	equilibrium when \(\mathbf D\) commutes with \(\mathbf A\).
	For example,
	\begin{equation}
		\mathbf A=\mathbf D=
		\begin{pmatrix}
			2&1\\
			1&3
		\end{pmatrix}.
		\label{eq:si-mode-equilibrium-noise}
	\end{equation}
	In the orthonormal eigenbasis of \(\mathbf A\),
	\(\boldsymbol{\Gamma}=\boldsymbol{\Lambda}\) is diagonal,
	so the equilibrium conditions are satisfied.
	
	\paragraph{Nonreciprocal normal networks.}
	For \(\mathbf A\neq\mathbf A^{\mathrm T}\) but
	\(\mathbf A\mathbf A^{\mathrm T}
	=\mathbf A^{\mathrm T}\mathbf A\),
	an orthonormal complex eigenbasis exists and at least
	one conjugate eigenvalue pair is nonreal.
	The corresponding ratios \(R_i(\tau)\) differ from unity
	for generic \(\tau>0\), excluding equilibrium for any
	\(\mathbf D\succ0\).
	For \(\mathbf D=\mathbf I\),
	\(\boldsymbol{\Gamma}=\mathbf I\), so distinct modes
	are uncorrelated and time-reversal asymmetry arises
	from fluctuations of individual modes.
	Spatially heterogeneous or correlated driving can
	additionally generate off-diagonal entries of
	\(\boldsymbol{\Gamma}=\mathbf U^\dagger\mathbf D\mathbf U\).
	For example,
	\begin{equation}
		\mathbf A=
		\begin{pmatrix}
			1&-1&0\\
			1&1&0\\
			0&0&2
		\end{pmatrix},
		\qquad
		\mathbf D=
		\begin{pmatrix}
			1&0&0.5\\
			0&1&0\\
			0.5&0&1
		\end{pmatrix}.
		\label{eq:si-mode-normal-example}
	\end{equation}
	In a suitable orthonormal eigenbasis,
	\begin{equation}
		\boldsymbol{\Lambda}=
		\begin{pmatrix}
			1+\mathrm i&0&0\\
			0&2&0\\
			0&0&1-\mathrm i
		\end{pmatrix},
		\qquad
		\boldsymbol{\Gamma}=
		\begin{pmatrix}
			1&\frac{0.5}{\sqrt{2}}&0\\
			\frac{0.5}{\sqrt{2}}&1&\frac{0.5}{\sqrt{2}}\\
			0&\frac{0.5}{\sqrt{2}}&1
		\end{pmatrix}.
		\label{eq:si-mode-normal-matrices}
	\end{equation}
	Here, \(R_1(\tau)=e^{-2\mathrm i\tau}\) and
	\(R_{12}(\tau)=e^{(1-\mathrm i)\tau}\), demonstrating
	contributions from both fluctuations of individual modes
	and correlated fluctuations between distinct modes.
	
	\paragraph{Nonnormal networks with an entirely real spectrum.}
	For
	\(\mathbf A\mathbf A^{\mathrm T}
	\neq\mathbf A^{\mathrm T}\mathbf A\)
	with \(\lambda_i\in\mathbb R\) for all \(i\),
	\(\mathbf U\) can be chosen real but not orthogonal.
	Thus, \(R_i(\tau)=1\), and any time-reversal asymmetry
	arises from correlated fluctuations between distinct modes.
	For \(\mathbf D=\mathbf I\),
	\(\boldsymbol{\Gamma}
	=(\mathbf U^{\mathrm T}\mathbf U)^{-1}\)
	has nonzero off-diagonal entries due to mode nonorthogonality.
	Whenever \(\Gamma_{ij}\neq0\) and
	\(\lambda_i\neq\lambda_j\),
	\(R_{ij}(\tau)\neq1\) for \(\tau>0\).
	For example,
	\begin{equation}
		\mathbf A=
		\begin{pmatrix}
			1&1\\
			0&2
		\end{pmatrix},
		\qquad
		\mathbf D=\mathbf I.
		\label{eq:si-mode-nonnormal-real-example}
	\end{equation}
	In a suitable real eigenbasis,
	\begin{equation}
		\boldsymbol{\Lambda}=
		\begin{pmatrix}
			1&0\\
			0&2
		\end{pmatrix},
		\qquad
		\boldsymbol{\Gamma}=
		\begin{pmatrix}
			2&-1\\
			-1&1
		\end{pmatrix}.
		\label{eq:si-mode-nonnormal-real-matrices}
	\end{equation}
	Here, \(R_{12}(\tau)=e^\tau\neq1\) for \(\tau>0\).
	Spatially correlated driving can restore equilibrium
	when \(\mathbf A\mathbf D=\mathbf D\mathbf A^{\mathrm T}\).
	For the same \(\mathbf A\), choosing
	\begin{equation}
		\mathbf D=
		\begin{pmatrix}
			2&1\\
			1&1
		\end{pmatrix}
		\label{eq:si-mode-nonnormal-equilibrium-noise}
	\end{equation}
	gives \(\boldsymbol{\Gamma}=\mathbf I\) in the same
	eigenbasis, satisfying the equilibrium conditions.
	
	\paragraph{Nonnormal networks with nonreal eigenvalues.}
	For
	\(\mathbf A\mathbf A^{\mathrm T}
	\neq\mathbf A^{\mathrm T}\mathbf A\)
	with at least one nonreal eigenvalue pair,
	the corresponding \(R_i(\tau)\neq1\) for generic
	\(\tau>0\) and any \(\mathbf D\succ0\).
	As in case~(iii), mode nonorthogonality and the spatial
	structure of stochastic driving jointly determine
	the off-diagonal entries of \(\boldsymbol{\Gamma}\).
	For example,
	\begin{equation}
		\mathbf A=
		\begin{pmatrix}
			1&-1&1\\
			1&1&0\\
			0&0&2
		\end{pmatrix},
		\qquad
		\mathbf D=\mathbf I.
		\label{eq:si-mode-mixed-example}
	\end{equation}
	In a suitable eigenbasis,
	\begin{equation}
		\boldsymbol{\Lambda}=
		\begin{pmatrix}
			1+\mathrm i&0&0\\
			0&2&0\\
			0&0&1-\mathrm i
		\end{pmatrix},
		\qquad
		\boldsymbol{\Gamma}=
		\begin{pmatrix}
			1.25&-0.5&0.25\\
			-0.5&1&-0.5\\
			0.25&-0.5&1.25
		\end{pmatrix}.
		\label{eq:si-mode-mixed-matrices}
	\end{equation}
	Here, time-reversal asymmetry arises from both fluctuations
	of individual modes and correlated fluctuations between
	distinct modes under homogeneous independent driving.
	
	\setcounter{equation}{0}
	\renewcommand{\theequation}{B\arabic{equation}}
	
	\section{Numerical validation of the spectral bounds}
	\label{sec:si-response-bound-validation}
	
\begin{figure*}[t]
		\centering
		\includegraphics{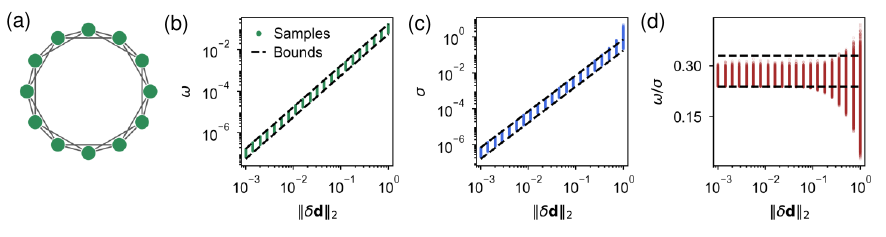}
		\caption{
			Numerical validation of the spectral bounds.
			(a) A binary regular ring with \(N=12\) nodes,
			each connected to its two nearest neighbors
			on either side.
			(b--d) Global asymmetry intensity \(\omega\),
			entropy production rate \(\sigma\), and their
			ratio \(\omega/\sigma\), respectively, evaluated
			from the exact steady-state expressions as functions
			of the driving heterogeneity \(\|\delta\mathbf d\|_2\).
			At each norm, dots represent \(10^5\) random
			driving patterns \(\mathbf d=\mathbf1+\delta\mathbf d\)
			with \(\delta\mathbf d\perp\mathbf1\) and \(d_m>0\).
			The dynamical parameters are \(\gamma=g=1\).
			Black dashed lines show the spectral predictions
			in Eqs.~(17) and (18) of the main text.
			The bounds on \(\omega\) are exact throughout
			the tested range, whereas those on \(\sigma\)
			and \(\omega/\sigma\) hold to leading order
			near equilibrium.
		}
		\label{fig:si-response-bound-validation}
	\end{figure*}

	We evaluate the spectral bounds in Eqs.~(17) and (18)
	of the main text on a binary regular ring with \(N=12\)
	nodes, each connected to its two nearest neighbors
	on either side
	[Fig.~\ref{fig:si-response-bound-validation}(a)].
	The interaction matrix is
	\(\mathbf A=\gamma\mathbf I+g\mathbf L\),
	with \(\gamma=g=1\).
	At each prescribed magnitude of driving heterogeneity,
	we sample \(10^5\) random patterns
	\(\mathbf d=\mathbf1+\delta\mathbf d\) satisfying
	\(\delta\mathbf d\perp\mathbf1\) and \(d_m>0\).
	The norm \(\|\delta\mathbf d\|_2\) is varied logarithmically
	from \(10^{-3}\) to \(1\), while the mean driving intensity
	remains fixed at unity.
	The global asymmetry intensity \(\omega\) and entropy
	production rate \(\sigma\) are evaluated from the exact
	expressions in Eq.~(12) of the main text.
	Figure~\ref{fig:si-response-bound-validation}(b--d)
	compares these quantities and their ratio with the
	spectral predictions.
	In panels (b) and (c), the dashed lines show
	\(\mu_{\min+}^{\alpha}\|\delta\mathbf d\|_2^2\)
	and \(\mu_{\max}^{\alpha}\|\delta\mathbf d\|_2^2\),
	with \(\alpha=\Omega\) and \(\mathrm J\), respectively.
	In panel (d), they show the lower and upper bounds
	of the generalized spectrum, \(\rho_{\min}\) and
	\(\rho_{\max}\).
	All sampled global asymmetry intensities lie within
	their spectral bounds throughout the tested range,
	consistent with the exact quadratic dependence of
	\(\omega\) on the driving intensities.
	For weak driving heterogeneity, the sampled entropy
	production rates and ratios \(\omega/\sigma\) also
	lie within their predicted intervals.
	As heterogeneity increases, some entropy production
	rates exceed the upper quadratic prediction, while
	the ratios extend beyond both generalized spectral bounds.
	This difference follows from the factors \(1/(d_md_n)\)
	in the exact entropy production rate.
	Near uniform driving, replacing these factors by unity
	gives the leading-order quadratic form
	\(\sigma_2=\delta\mathbf d^{\mathrm T}
	\boldsymbol{\Theta}^{\mathrm J}\delta\mathbf d\)
	in Eq.~(15) of the main text.
	The response and generalized spectra bound
	\(\sigma_2\) and \(\omega/\sigma_2\) exactly.
	Their application to the exact entropy production rate
	and its ratio with \(\omega\) is therefore restricted
	to the near-equilibrium regime, where higher-order
	contributions in the driving heterogeneity are small.
	
	\setcounter{equation}{0}
	\renewcommand{\theequation}{C\arabic{equation}}
	
	\section{Generation and spectral bounds of hierarchical modular networks}
	\label{sec:si-network-generation}
	
	Throughout this section, the response spectral bounds
	refer to \(\mu_{\min+}^{\alpha}\) and
	\(\mu_{\max}^{\alpha}\), with
	\(\alpha=\Omega,\mathrm{J}\).
	The generalized and nonzero Laplacian spectral bounds
	are \((\rho_{\min},\rho_{\max})\) and
	\((k_2,k_N)\), respectively.
	
	\subsection{Binary synthetic hierarchical modular networks}
	
	\paragraph{Network generation.}
	We generate undirected binary networks without self-connections.
	For each parameter setting, the number of nodes \(N\)
	and connections \(E=N\langle e\rangle/2\) is fixed
	across hierarchical classes.
	The finest partition contains \(M_1=2^H\) modules,
	each with \(n=N/M_1\) nodes, which are recursively
	paired to form larger modules.
	Node pairs are classified by the smallest hierarchical
	group containing both nodes.
	The corresponding numbers of candidate neighbors per node
	are \(c_1=n-1\) and \(c_\ell=2^{\ell-2}n\) for
	\(\ell=2,\ldots,H+1\).
	The target connection probabilities are
	\begin{equation}
		p_\ell=p_1r^{\ell-1},
		\quad
		p_1=\frac{\langle e\rangle}
		{\sum_{\ell=1}^{H+1}c_\ell r^{\ell-1}},
		\quad 0<r<1,
		\label{eq:si-hmn-probabilities}
	\end{equation}
	with parameters chosen such that \(p_1\leq1\).
	The unrounded connection count in category \(\ell\)
	is \(\widetilde E_\ell=Nc_\ell p_\ell/2\).
	We initially set \(E_\ell=\lfloor\widetilde E_\ell\rfloor\)
	and allocate the remaining connections to categories
	with the largest fractional remainders, ensuring
	\(\sum_\ell E_\ell=E\).
	The class \(\mathrm{H}_h\), with \(h=0,\ldots,H\),
	retains categories \(1,\ldots,h\) and merges
	the remaining categories.
	The merged target probability and connection count are
	\begin{equation}
		p_{\mathrm{out}}^{(h)}
		=\frac{\sum_{\ell=h+1}^{H+1}c_\ell p_\ell}
		{\sum_{\ell=h+1}^{H+1}c_\ell},
		\qquad
		E_{\mathrm{out}}^{(h)}
		=\sum_{\ell=h+1}^{H+1}E_\ell.
		\label{eq:si-hmn-level-reduction}
	\end{equation}
	For \(M_1=8\), \(\mathrm{H}_3\) contains three nested
	levels with \(8\), \(4\), and \(2\) modules.
	The classes \(\mathrm{H}_2\) and \(\mathrm{H}_1\)
	retain the levels with \(8\) and \(4\) modules,
	and with \(8\) modules, respectively.
	In \(\mathrm{H}_0\), all node pairs belong to one category,
	with \(p_{\mathrm{out}}^{(0)}=\langle e\rangle/(N-1)\).
	Within each retained or merged category, the prescribed
	number of unordered node pairs is sampled uniformly
	without replacement.
	Classes are sampled independently using the same
	category counts \(E_\ell\).
	Thus, every realization has exactly \(E\) connections,
	with \(\mathrm{H}_0\) providing the homogeneous random
	reference.
	
	\paragraph{Parameter settings.}
	The baseline networks have \(N=360\),
	\(\langle e\rangle=60\), \(M_1=8\), and \(r=0.5\).
	The dynamics follow
	\(\mathbf A=\gamma\mathbf I+g\mathbf L\),
	with baseline parameters \(\gamma=g=1\) and
	relaxation rates \(\lambda_i=\gamma+gk_i\).
	We generate \(100\) realizations for each network class
	and parameter setting.
	One parameter is varied at a time.
	Network size is varied over \(N=360,720,1440\)
	at fixed \(\langle e\rangle=60\), and mean degree
	over \(\langle e\rangle=30,60,90\) at fixed \(N=360\).
	The parameters \(M_1=8\) and \(r=0.5\) are retained
	throughout.
	Eigenvalues are sorted within each realization and
	averaged at each rank.
	Uniform zero modes are omitted from the response
	operator and Laplacian spectra.
	
\begin{figure*}[t]
		\centering
		\includegraphics{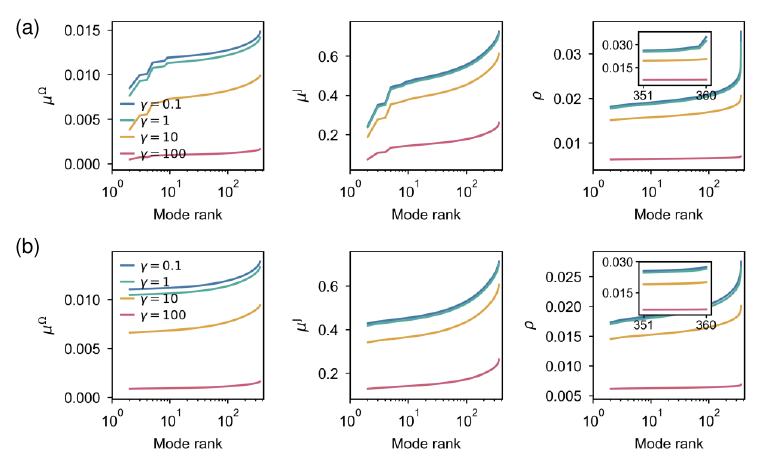}
		\caption{
			Effect of the local relaxation rate \(\gamma\)
			on spectral bounds.
			Rows (a) and (b) show \(\mathrm{H}_3\) and
			\(\mathrm{H}_0\), respectively.
			Columns show \(\{\mu_r^\Omega\}\),
			\(\{\mu_r^{\mathrm{J}}\}\), and \(\{\rho_r\}\)
			from left to right.
			Colors indicate \(\gamma=0.1,1,10,100\),
			with \(g=1\) and baseline binary network parameters.
		}
		\label{fig:si-gamma}
	\end{figure*}

	\paragraph{Effect of the local relaxation rate.}
	Figure~\ref{fig:si-gamma} shows the spectra for
	\(\mathrm{H}_3\) and \(\mathrm{H}_0\) as \(\gamma\)
	varies at fixed \(g=1\).
	Increasing \(\gamma\) raises all relaxation rates
	while preserving the relaxation modes and rate differences.
	The larger sums \(\lambda_i+\lambda_j\) reduce
	\(\kappa_{ij}^{\Omega}\) and
	\(\kappa_{ij}^{\mathrm{J}}\), lowering both bounds
	of the response operator spectra.
	The generalized spectrum also shifts downward,
	reducing its lower and upper bounds.
	
\begin{figure*}[t]
		\centering
		\includegraphics{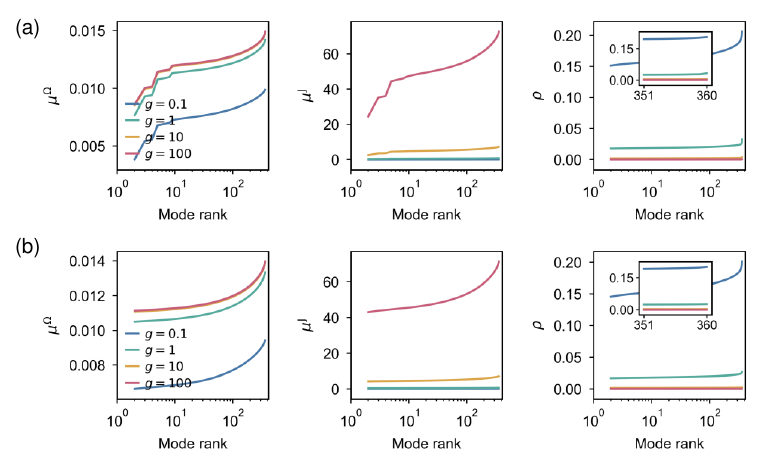}
		\caption{
			Effect of the coupling strength \(g\)
			on spectral bounds.
			Rows (a) and (b) show \(\mathrm{H}_3\) and
			\(\mathrm{H}_0\), respectively.
			Columns show \(\{\mu_r^\Omega\}\),
			\(\{\mu_r^{\mathrm{J}}\}\), and \(\{\rho_r\}\)
			from left to right.
			Colors indicate \(g=0.1,1,10,100\),
			with \(\gamma=1\) and baseline binary network parameters.
		}
		\label{fig:si-coupling}
	\end{figure*}

	\paragraph{Effect of the coupling strength.}
	Figure~\ref{fig:si-coupling} shows the spectra for
	\(\mathrm{H}_3\) and \(\mathrm{H}_0\) as \(g\)
	varies at fixed \(\gamma=1\).
	Increasing \(g\) enlarges the relaxation rate differences
	\(g(k_i-k_j)\), increasing both
	\(\kappa_{ij}^{\Omega}\) and
	\(\kappa_{ij}^{\mathrm{J}}\) and raising both response
	spectral bounds.
	At strong coupling, the bounds of
	\(\{\mu_r^\Omega\}\) approach finite limits because
	the squared rate differences and squared rate sums
	scale equally with \(g\).
	The bounds of \(\{\mu_r^{\mathrm{J}}\}\) instead
	grow approximately linearly with \(g\).
	The generalized spectral bounds decrease and
	scale approximately as \(1/g\) at strong coupling.
	
\begin{figure*}[t]
		\centering
		\includegraphics{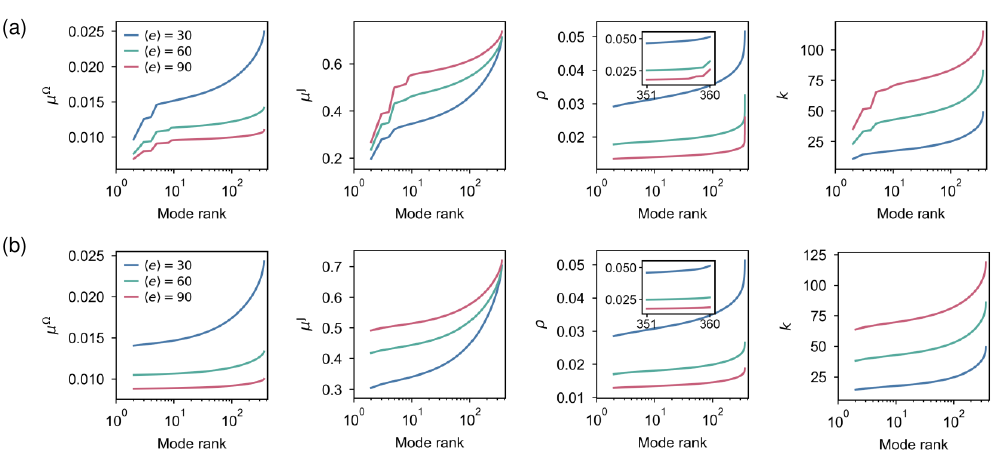}
		\caption{
			Effect of the mean degree \(\langle e\rangle\)
			on spectral bounds.
			Rows (a) and (b) show \(\mathrm{H}_3\) and
			\(\mathrm{H}_0\), respectively.
			Columns show \(\{\mu_r^\Omega\}\),
			\(\{\mu_r^{\mathrm{J}}\}\), \(\{\rho_r\}\),
			and \(\{k_i\}\) from left to right.
			Colors indicate \(\langle e\rangle=30,60,90\),
			with \(N=360\), \(\gamma=g=1\), and the remaining
			network parameters fixed at their baseline values.
		}
		\label{fig:si-degree}
	\end{figure*}

	\paragraph{Effect of the mean degree.}
	Figure~\ref{fig:si-degree} shows the spectra for
	\(\mathrm{H}_3\) and \(\mathrm{H}_0\) as
	\(\langle e\rangle\) varies at fixed \(N=360\)
	and \(\gamma=g=1\).
	Increasing \(\langle e\rangle\) raises the target
	connection probabilities while preserving their ratios
	across hierarchical categories.
	The Laplacian spectrum shifts upward, consistent
	with the increasing spectral sum
	\(\sum_i k_i=N\langle e\rangle\).
	Over the tested range, both bounds of
	\(\{\mu_r^{\mathrm{J}}\}\) increase, whereas those of
	\(\{\mu_r^\Omega\}\) and \(\{\rho_r\}\) decrease.
	The decrease in the bounds of \(\{\mu_r^\Omega\}\)
	differs from their increase under stronger coupling:
	changing connectivity alters both the Laplacian
	eigenvalues and modes, whereas varying \(g\)
	rescales the existing couplings.
	The differences between \(\mathrm{H}_3\) and
	\(\mathrm{H}_0\) become more pronounced at higher
	mean degree.
	
\begin{figure*}[t]
		\centering
		\includegraphics{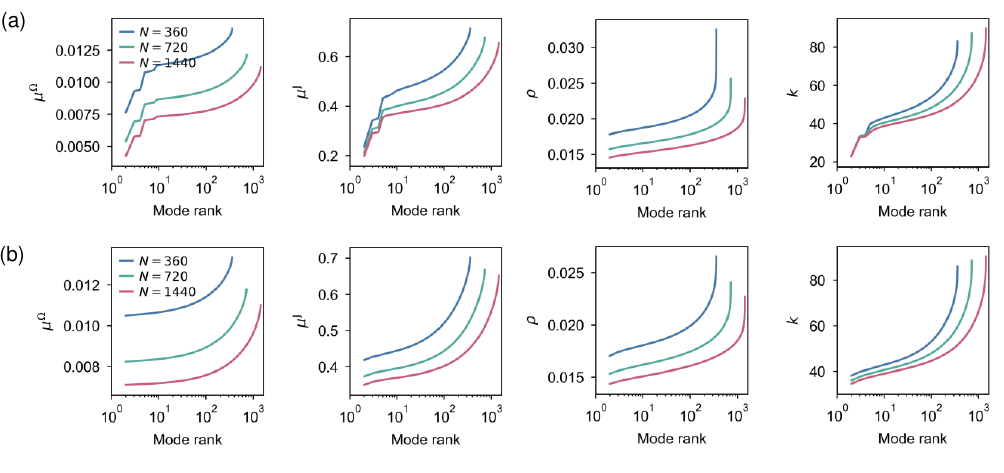}
		\caption{
			Effect of network size \(N\) on spectral bounds.
			Rows (a) and (b) show \(\mathrm{H}_3\) and
			\(\mathrm{H}_0\), respectively.
			Columns show \(\{\mu_r^\Omega\}\),
			\(\{\mu_r^{\mathrm{J}}\}\), \(\{\rho_r\}\),
			and \(\{k_i\}\) from left to right.
			Colors indicate \(N=360,720,1440\),
			with \(\langle e\rangle=60\), \(\gamma=g=1\),
			and the remaining network parameters fixed
			at their baseline values.
		}
		\label{fig:si-size}
	\end{figure*}

	\paragraph{Effect of network size.}
	Figure~\ref{fig:si-size} shows the spectra for
	\(\mathrm{H}_3\) and \(\mathrm{H}_0\) as \(N\)
	increases at fixed \(\langle e\rangle=60\)
	and \(\gamma=g=1\).
	With \(M_1=8\) fixed, increasing \(N\) enlarges
	each module and reduces the connection density.
	The Laplacian spectrum broadens modestly through
	an increase in its upper bound \(k_N\), while
	its mean nonzero eigenvalue,
	\(N\langle e\rangle/(N-1)\), remains nearly constant.
	Both bounds of \(\{\mu_r^\Omega\}\) decrease,
	whereas those of \(\{\mu_r^{\mathrm{J}}\}\)
	show a more modest decrease.
	The generalized spectral bounds also decrease,
	particularly \(\rho_{\max}\) in \(\mathrm{H}_3\),
	reducing its difference from \(\mathrm{H}_0\).
	Across these parameter scans, \(\mathrm{H}_3\)
	retains smaller lower response bounds
	\(\mu_{\min+}^{\Omega}\) and
	\(\mu_{\min+}^{\mathrm{J}}\) and a larger upper
	generalized bound \(\rho_{\max}\) than
	\(\mathrm{H}_0\).
	
	\subsection{Weighted synthetic hierarchical modular networks}
	
	\paragraph{Network generation.}
	Weighted networks are constructed by assigning positive
	weights to the existing connections of each binary
	realization.
	For each realization, we draw independent samples
	\(z_e\sim\mathcal N(0,1)\), \(e=1,\ldots,E\), and set
	\begin{equation}
		W_e=E\,
		\frac{\exp\!\left[z_e\sqrt{\ln(1+\mathrm{CV}^2)}\right]}
		{\sum_{q=1}^{E}
			\exp\!\left[z_q\sqrt{\ln(1+\mathrm{CV}^2)}\right]}.
		\label{eq:si-normalized-connection-weights}
	\end{equation}
	The same Gaussian samples are used across hierarchical
	classes and CV values within each realization.
	For each class, weights are randomly assigned to
	existing connections, with the assignment held fixed
	as CV varies.
	Weights are assigned symmetrically, preserving
	\(W_{ij}=W_{ji}\).
	Normalization fixes the mean connection weight at unity
	and the total connection weight at
	\(\sum_{e=1}^{E}W_e=E\).
	Matched hierarchical classes therefore share the same
	weight set, while varying CV changes weight heterogeneity
	at fixed binary connectivity and total weight.
	The binary network is recovered at \(\mathrm{CV}=0\).
	For a set of positive connection weights, define
	\(W_m=E^{-1}\sum_{e=1}^{E}W_e\).
	The empirical coefficient of variation is
	\begin{equation}
		\mathrm{CV}_{\mathrm{emp}}
		=\frac{\sqrt{E^{-1}\sum_{e=1}^{E}(W_e-W_m)^2}}
		{W_m}.
		\label{eq:si-empirical-weight-cv}
	\end{equation}
	Here \(W_m=1\), but finite sampling can produce
	differences between \(\mathrm{CV}_{\mathrm{emp}}\)
	and the prescribed CV of the underlying lognormal
	distribution.
	
	\paragraph{Parameter settings.}
	We vary \(\mathrm{CV}=0,0.5,1,1.5,2\) at
	\(N=360\), \(\langle e\rangle=60\),
	\(M_1=8\), \(r=0.5\), and \(\gamma=g=1\).
	Each network class contains \(100\) realizations.
	Spectra are sorted within each realization and
	averaged at each rank, with uniform zero modes omitted
	from the response operator and Laplacian spectra.
	
\begin{figure*}[t]
		\centering
		\includegraphics{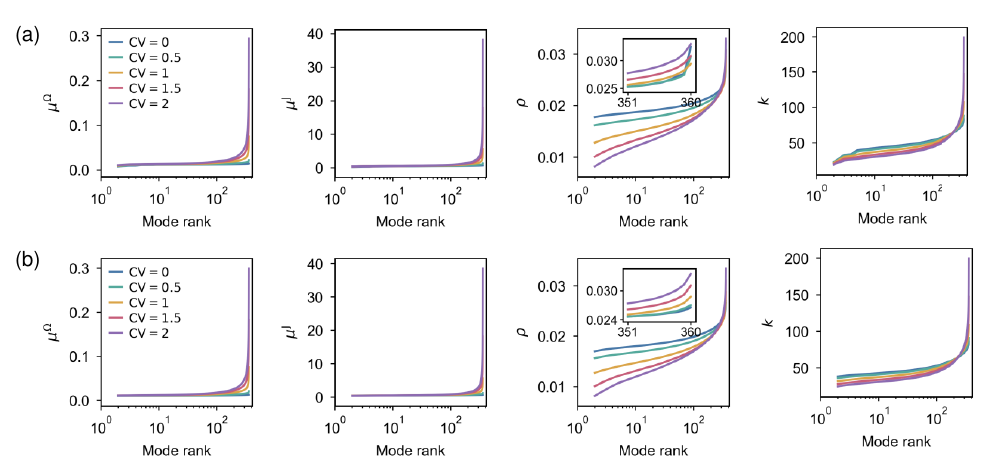}
		\caption{
			Effect of weight heterogeneity on spectral bounds.
			Rows (a) and (b) show \(\mathrm{H}_3\) and
			\(\mathrm{H}_0\), respectively.
			Columns show \(\{\mu_r^\Omega\}\),
			\(\{\mu_r^{\mathrm{J}}\}\), \(\{\rho_r\}\),
			and \(\{k_i\}\) from left to right.
			Colors indicate \(\mathrm{CV}=0,0.5,1,1.5,2\),
			with \(\gamma=g=1\) and baseline network parameters.
			Binary connectivity and total connection weight
			are fixed as CV varies.
		}
		\label{fig:si-heterogeneity}
	\end{figure*}

	\paragraph{Effect of weight heterogeneity.}
	Figure~\ref{fig:si-heterogeneity} shows the spectra for
	\(\mathrm{H}_3\) and \(\mathrm{H}_0\) as CV increases.
	Both response operator spectra broaden primarily through
	marked increases in their upper bounds
	\(\mu_{\max}^{\Omega}\) and
	\(\mu_{\max}^{\mathrm{J}}\), whereas changes in
	their lower bounds are comparatively smaller.
	The upper Laplacian bound \(k_N\) also increases,
	introducing faster relaxation rates while preserving
	the spectral sum \(\sum_i k_i=2E\).
	Meanwhile, the lower generalized bound
	\(\rho_{\min}\) decreases.
	This trend is consistent with
	\(\kappa_{ij}^{\Omega}/\kappa_{ij}^{\mathrm{J}}
	=2/(\lambda_i+\lambda_j)\) for contributing mode pairs,
	since larger relaxation rate sums give smaller ratios.
	These effects occur in both hierarchical and random
	networks.
	
	\subsection{HMR and HMH in structural connectivity networks}
	
	Surrogate networks are constructed separately for each
	participant from the weighted, undirected SC matrix
	\(\mathbf W\), with \(N=360\) and \(W_{ii}=0\).
	The weights of its \(E\) existing connections are
	denoted by \(W_e\), counting each undirected connection once.
	Both procedures preserve the number of nodes and
	connections and the total connection weight of each
	participant, without additional normalization
	or thresholding.
	
	\paragraph{Hierarchical modular randomization.}
	The nested partitions are determined from the empirical
	Laplacian modes ordered by increasing eigenvalue.
	The signs of \(\mathbf U_2\) divide the network into
	two modules, and those of \(\mathbf U_3\) subdivide
	each module, giving up to four finer modules.
	These partitions are held fixed during randomization.
	Within each selected module, unordered node pairs
	are sampled uniformly without replacement from all
	possible pairs, keeping their number equal to the
	empirical connection count.
	The original nonzero weights are randomly assigned
	to these pairs, preserving the connection count and
	weight set within the module.
	Connections between selected modules remain unchanged.
	Randomization is applied within the finer modules
	for \(\mathrm{H}_2\), within the two largest modules
	for \(\mathrm{H}_1\), and across the whole network
	for \(\mathrm{H}_0\).
	Each condition is generated directly from empirical SC,
	with one independent realization per participant.
	Node degrees and strengths are not constrained.
	
	\paragraph{Hierarchical modular homogenization.}
	Connection positions are fixed, and existing weights
	are transformed as
	\begin{equation}
		W_e^{(\upsilon)}
		=\upsilon W_e+(1-\upsilon)W_m,
		\qquad 0\leq\upsilon\leq1,
		\label{eq:si-hmh-weights}
	\end{equation}
	where \(W_m=E^{-1}\sum_{e=1}^{E}W_e\) is the mean
	weight of existing connections.
	Absent connections remain zero, and weights are
	assigned symmetrically.
	Each value of \(\upsilon\) is applied directly to
	the empirical weights.
	The empirical SC is recovered at \(\upsilon=1\),
	whereas \(\upsilon=0\) assigns \(W_m\) to every
	existing connection.
	Using Eq.~\eqref{eq:si-empirical-weight-cv},
	the coefficient of variation satisfies
	\begin{equation}
		\mathrm{CV}(\upsilon)
		=\upsilon\,\mathrm{CV}_{\mathrm{emp}},
		\label{eq:si-hmh-cv}
	\end{equation}
	where \(\mathrm{CV}_{\mathrm{emp}}\) is evaluated
	for the empirical SC weights.
	
\begin{figure*}[t]
		\centering
		\includegraphics{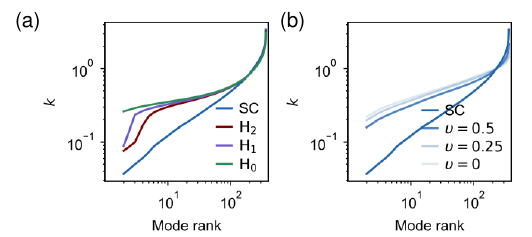}
		\caption{
			Laplacian spectra of human structural connectomes
			and surrogate networks.
			(a) HMR, with blue, red, purple, and green curves
			denoting SC, \(\mathrm{H}_2\), \(\mathrm{H}_1\),
			and \(\mathrm{H}_0\), respectively.
			(b) HMH, with progressively lighter blue curves
			denoting \(\upsilon=1\) (SC), \(0.5\), \(0.25\),
			and \(0\).
			All networks contain \(N=360\) cortical regions.
			Eigenvalues are sorted within each participant
			and averaged at each rank over \(993\) participants.
			The uniform zero mode is omitted.
		}
		\label{fig:si-brain-laplacian}
	\end{figure*}

	\paragraph{Laplacian spectra in HMR and HMH.}
	Figure~\ref{fig:si-brain-laplacian} compares the Laplacian
	spectra of empirical SC and its surrogates.
	In HMR, randomization over progressively coarser regions
	raises the lower bound \(k_2\), while the upper bound
	\(k_N\) remains comparatively unchanged.
	In HMH, decreasing \(\upsilon\) raises \(k_2\) and
	lowers \(k_N\), with complete weight homogenization
	(\(\upsilon=0\)) producing the narrowest spectrum.
	Both procedures preserve the spectral sum
	\(\sum_i k_i=2EW_m\).
	At fixed \(\gamma\) and \(g\), these changes narrow
	the range of nonuniform relaxation rates
	\(\lambda_i=\gamma+gk_i\), \(i\geq2\).
	
	\begin{figure*}[t]
		\centering
		\includegraphics{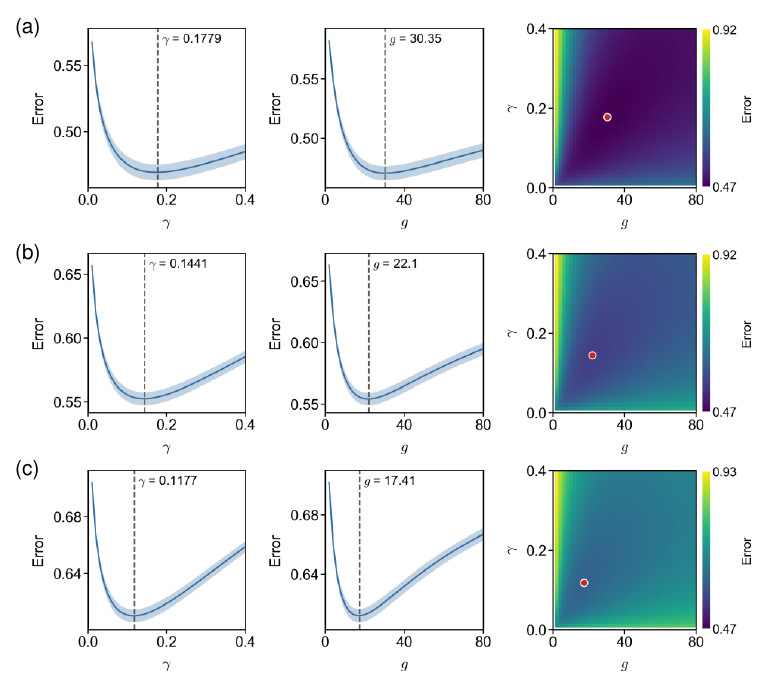}
		\caption{
			Parameter optimization for different sets of fitted lags.
			Rows (a--c) correspond to \(\mathcal T_{01}\),
			\(\mathcal T_{012}\), and \(\mathcal T_{0123}\),
			respectively.
			Left and middle columns show mean fitting-error
			profiles for \(\gamma\) and \(g\), obtained by
			minimizing over the other parameter within each
			participant before averaging across participants.
			Shading indicates pointwise \(95\%\) confidence
			intervals for the means.
			Right panels show participant-averaged error
			landscapes with a common color scale.
			Dashed lines and red markers indicate the means
			of individual optimal parameter estimates.
		}
		\label{fig:si-fitting-errors}
	\end{figure*}
	
	\begin{figure*}[t]
		\centering
		\includegraphics{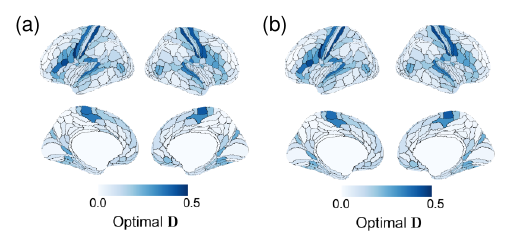}
		\caption{
			Optimal stochastic driving patterns inferred
			using additional lags.
			Panels (a) and (b) show group-averaged patterns
			for \(\mathcal T_{012}\) and \(\mathcal T_{0123}\),
			respectively.
			Regional driving intensities are min--max normalized
			to \([0,1]\) within each participant before averaging.
			Top and bottom rows show lateral and medial views,
			respectively.
			Both panels use the same color scale.
		}
		\label{fig:si-driving-additional-lags}
	\end{figure*}
	
	\begin{figure*}[t]
		\centering
		\includegraphics{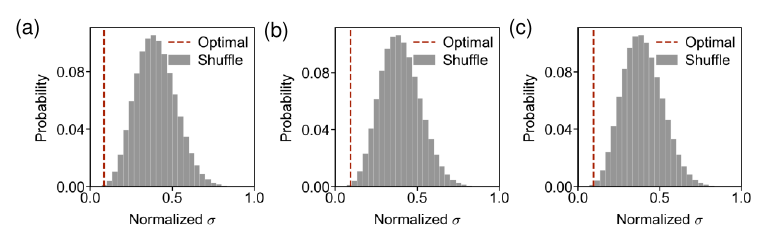}
		\caption{
			Entropy production rates for optimal and spatially shuffled
			driving patterns.
			Panels (a--c) correspond to \(\mathcal T_{01}\),
			\(\mathcal T_{012}\), and \(\mathcal T_{0123}\), respectively.
			The optimal and \(10^5\) shuffled rates are jointly min--max
			normalized to \([0,1]\) within each participant.
			Gray histograms show the fraction of all normalized shuffled
			rates in each of \(30\) equal bins.
			Red dashed lines mark the mean normalized \(\sigma\) of the
			optimal patterns across participants.
		}
		\label{fig:si-epr-shuffle}
	\end{figure*}
	
	\begin{figure*}[t]
		\centering
		\includegraphics{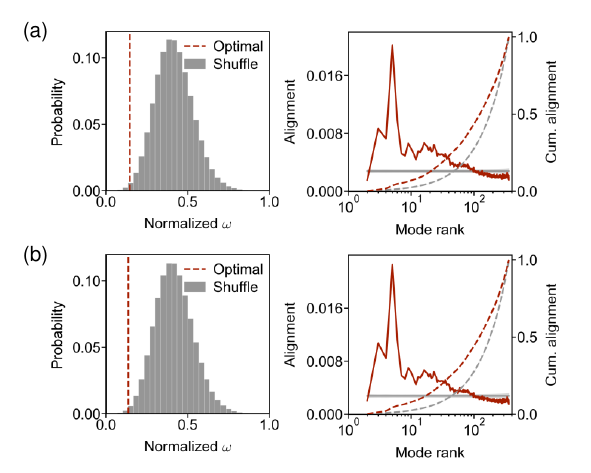}
		\caption{
			Global asymmetry intensity and driving alignment for fits
			including additional lags.
			Rows (a) and (b) correspond to \(\mathcal T_{012}\) and
			\(\mathcal T_{0123}\), respectively.
			Left panels show \(\omega\) after joint min--max normalization
			of the optimal and \(10^5\) shuffled intensities within each
			participant.
			Gray histograms pool the normalized shuffled intensities
			across participants, and red dashed lines mark the mean
			for the optimal patterns.
			Right panels show alignment of optimal (red) and shuffled
			(gray) patterns with the nonuniform modes of
			\(\boldsymbol{\Theta}^{\Omega}\), ordered by increasing
			eigenvalue within each participant.
			Solid curves show mean alignment strength (left axis).
			Dashed curves show mean cumulative alignment (right axis).
			Gray shading indicates pointwise central \(95\%\) intervals
			of the group shuffle distributions.
		}
		\label{fig:si-shuffle-additional-lags}
	\end{figure*}

	\setcounter{equation}{0}
	\renewcommand{\theequation}{D\arabic{equation}}
	
	\section{Participants and data processing}
	
	\subsection{Data acquisition}
	
	The dataset was obtained from the WU--Minn Human Connectome
	Project (HCP) Young Adult database~\cite{VanEssen2013} and
	included structural magnetic resonance imaging (MRI),
	diffusion-weighted imaging (DWI), and resting-state functional
	MRI (fMRI) data from 993 healthy young adults
	(age range, 22--36 years).
	Each participant completed four resting-state fMRI runs
	acquired across two sessions, with \(1{,}200\) time points
	per run and a repetition time of
	\(\mathrm{TR}=0.72\)~s~\cite{Smith2013HCP},
	yielding \(4{,}800\) time points in total.
	
	\subsection{Resting-state fMRI data processing}
	
	We used resting-state fMRI data preprocessed by the HCP
	minimal preprocessing pipelines~\cite{Glasser2013}.
	Preprocessing included correction for gradient nonlinearities,
	head motion, and susceptibility-induced distortions, followed
	by registration to structural images, alignment to standard
	space, and projection onto the cortical surface.
	The preprocessed data were provided as grayordinate time
	series in CIFTI format.
	The cerebral cortex was parcellated into 360 regions
	(180 per hemisphere) using the HCP multimodal parcellation
	atlas~\cite{Glasser2016}.
	
	For each run, regional blood-oxygen-level-dependent (BOLD)
	time series were obtained by averaging cortical grayordinate
	signals within each region and then band-pass filtered
	between 0.01 and 0.1~Hz to reduce slow drifts and
	high-frequency fluctuations.
	Global signal regression was not performed.
	Temporal processing was conducted separately for each run
	before concatenating the four runs in acquisition order,
	yielding \(4{,}800\) time points per region for subsequent
	analyses.
	
	\subsection{Diffusion MRI data processing}
	
	The DWI data were preprocessed using the HCP minimal
	preprocessing pipeline~\cite{Glasser2013}, including correction
	for gradient nonlinearities, subject motion, eddy-current
	and susceptibility-induced distortions, followed by registration
	to the corresponding T1-weighted image.
	Further processing was performed in MRtrix3~\cite{Tournier2019}.
	Response functions for white matter, gray matter, and
	cerebrospinal fluid were estimated using the
	\textit{dhollander} algorithm~\cite{Dhollander2019},
	followed by multi-shell, multi-tissue constrained spherical
	deconvolution~\cite{Jeurissen2014}.
	Anatomically constrained tractography used a five-tissue-type
	image generated from the T1-weighted image, with backtracking
	enabled~\cite{Smith2012ACT}.
	Probabilistic streamlines were generated using the iFOD2
	algorithm~\cite{Tournier2010}, with dynamic seeding based on
	the white-matter fiber-orientation distribution.
	Ten million streamlines were generated per participant
	and subsequently weighted using SIFT2~\cite{Smith2015SIFT2}.
	Streamline endpoints were assigned to the 360 cortical
	regions defined by the same multimodal parcellation atlas.
	The structural connectivity weight between regions
	\(i\) and \(j\) was defined as
	\(W_{ij}=\sum_{s\in\mathcal S_{ij}}\mu_s\),
	where \(\mathcal S_{ij}\) contains the streamlines connecting
	the two regions and \(\mu_s\) is the SIFT2 weight of
	streamline \(s\).
	This yielded a nonnegative, weighted, and undirected
	\(360\times360\) structural connectivity matrix for each
	participant, with self-connections set to zero.
    
    \setcounter{equation}{0}
    \renewcommand{\theequation}{E\arabic{equation}}
    
    \section{Optimization settings and robustness of inferred stochastic driving}
    \label{sec:si-driving-inference}
    
    \subsection{Parameter optimization and fitting errors}
    
    The inference procedure in Appendix H of the main text
    is applied separately to \(993\) participants, each with
    \(N=360\) brain regions.
    The search grid comprises
    \(\gamma=0.01,0.02,\ldots,0.40\) and
    \(g=2,4,\ldots,80\), giving \(1600\) candidate pairs.
    At each pair, the driving intensities are optimized using
    the \texttt{interior-point-convex} algorithm in MATLAB's
    \texttt{quadprog}, subject to \(d_i\geq10^{-6}\).
    The parameter pair minimizing the mean squared relative
    Frobenius error across the fitted targets, together with
    its optimized driving intensities, defines the optimal fit.
    The main analysis uses \(\mathcal T_{01}=\{0,1\}\),
    jointly fitting the zero-lag covariance and one-lag
    cross-correlation asymmetry.
    Additional fits use \(\mathcal T_{012}=\{0,1,2\}\) and
    \(\mathcal T_{0123}=\{0,1,2,3\}\), with lags measured
    in sampling intervals (\(\mathrm{TR}=0.72\,\mathrm{s}\)).
    All parameters and driving intensities are reoptimized
    for each fitting scheme using the same grid and constraints.
    Figure~\ref{fig:si-fitting-errors} summarizes the fitting
    errors and parameter estimates.
    Group summaries include fits with finite errors throughout
    the grid and finite optimal parameter estimates.
    
    \subsection{Spatial driving patterns inferred using additional lags}
    
    The group-averaged driving patterns inferred using
    \(\mathcal T_{012}\) and \(\mathcal T_{0123}\)
    [Fig.~\ref{fig:si-driving-additional-lags}]
    retain the principal spatial features of the
    \(\mathcal T_{01}\) fit shown in Fig.~4(b) of the main text.
    Participant-wise min--max normalization is used for
    visualization, while the response analyses use the
    original fitted driving intensities.
    
    \subsection{Entropy production under spatial shuffling}
    
    For each participant and fitting scheme, we construct
    \(\boldsymbol{\Theta}^{\mathrm J}\) from the individual structural
    connectome using the optimal \(\gamma\) and \(g\).
    Holding the operator fixed, we evaluate the exact entropy production
    rate at steady state,
    \(\sigma=-\sum_m(\boldsymbol{\Theta}^{\mathrm J}\mathbf d)_m/d_m\),
    for the optimal driving pattern and for \(10^5\) spatial permutations
    of its regional intensities.
    The permutations are drawn independently for each participant
    and preserve the distribution of driving intensities.
    Within each participant and fitting scheme, we jointly min--max
    normalize the optimal and shuffled rates to \([0,1]\).
    For each permutation index, we average the normalized shuffled
    rates across participants to obtain a group mean.
    In each of the three fitting schemes, the mean normalized
    \(\sigma\) for the optimal patterns falls below all \(10^5\)
    shuffled group means (permutation test, \(p<10^{-4}\))
    [Fig.~\ref{fig:si-epr-shuffle}].
    
    \subsection{Global asymmetry intensity and mode alignment for fits including additional lags}
    
    Using the same spatial permutations, we evaluate
    \(\omega=\mathbf d^{\mathrm T}\boldsymbol{\Theta}^{\Omega}\mathbf d\)
    with \(\boldsymbol{\Theta}^{\Omega}\) held fixed for each participant
    and fitting scheme.
    Normalization and group permutation testing follow the procedure
    described above for \(\sigma\).
    We calculate alignment strengths as in the main text.
    For each permutation index, we average alignment strengths across
    participants at matching mode ranks and sum over successive ranks
    to obtain the cumulative alignment curve.
    The \(10^5\) group curves give the mean and pointwise central
    \(95\%\) intervals for each quantity.
    For both additional-lag fits, the optimal patterns yield lower
    mean normalized \(\omega\) than the shuffled patterns and align
    more strongly with modes associated with smaller eigenvalues.
    Their cumulative alignment also rises more rapidly at low mode
    ranks [Fig.~\ref{fig:si-shuffle-additional-lags}], consistent
    with Fig.~4(c,d) of the main text.
    \bibliography{references}